\documentclass[journal]{IEEEtran}
\ifCLASSINFOpdf
\else
\fi
\usepackage{amsmath,amssymb,amsfonts}
\usepackage{graphicx}
\usepackage{tkz-euclide}
\usepackage{siunitx}
\usepackage[percent]{overpic}
\usepackage{subcaption}
\usepackage{mathrsfs}
\usepackage[mathscr]{eucal}
\usepackage{pgfplots}
\pgfplotsset{compat=1.10}
\usepgfplotslibrary{fillbetween}
\usepackage{xfrac}
\usepackage{commath}
\usepackage{cite}
\usepackage{authblk}
\usepackage{multirow}
\usepackage{color, colortbl}
\definecolor{g}{rgb}{0.77, 0.76, 0.82}
\usepackage{mathabx}

\begin{document}
\bstctlcite{IEEEexample:BSTcontrol}

\title{Minimax Robust Landmine Detection Using Forward-Looking Ground-Penetrating Radar}

\author{A.~D.~Pambudi,~\IEEEmembership{Student Member,~IEEE,} M.~Fau\ss{},~\IEEEmembership{Member,~IEEE,} F.~Ahmad,~\IEEEmembership{Fellow,~IEEE,} and~A.~M.~Zoubir,~\IEEEmembership{Fellow,~IEEE}

\thanks{©2020 IEEE. Personal use of this material is permitted. Permission from IEEE  must  be  obtained  for  all  other  uses,  in  any  current  or  future media, including reprinting/republishing this material for advertising or promotional purposes, creating new collective works, for resale or redistribution to servers or lists, or reuse of any copyrighted component of this work in other works.}

\thanks{The work of A.~D.~Pambudi is supported by the `Excellence Initiative' of the German Federal and State Governments and the Graduate School of Computational Engineering at Technische Universit{\"a}t Darmstadt. He also is supported by Deutsche Akademische Austauschdienst (DAAD)-Indonesian German Scholarship Programme.}

\thanks{A.~D.~Pambudi is with the Signal Processing Group, Department of Electrical Engineering and Information Technology, Technische Universit{\"a}t Darmstadt, Darmstadt 64283, Germany, and also with the School of Electrical Engineering, Telkom University, Bandung 40257, Indonesia (on extended leave) (e-mail: apambudi@spg.tu-darmstadt.de; apambudi@telkomuniversity.ac.id).}

\thanks{M.~Fau\ss{},~and~A.~M.~Zoubir are with the Signal Processing Group, Department of Electrical Engineering and Information Technology, Technische Universit{\"a}t Darmstadt, Darmstadt 64283, Germany (e-mail: \{mfauss, zoubir\}@spg.tu-darmstadt.de).}

\thanks{F.~Ahmad is with the Department of Electrical and Computer Engineering, Temple University, Philadelphia, PA 19122, USA (e-mail: fauzia.ahmad@temple.edu).}}

\markboth{}%
{Shell \MakeLowercase{\textit{et al.}}: Bare Demo of IEEEtran.cls for IEEE Journals}

\maketitle

\begin{abstract}
We propose a robust likelihood-ratio test (LRT) to detect landmines and unexploded ordnance using a forward-looking ground-penetrating radar. Instead of modeling the distributions of the target and clutter returns with parametric families, we construct a band of feasible probability densities under each hypothesis. The LRT is then devised based on the least favorable densities within the bands. This detector is designed to maximize the worst-case performance over all feasible density pairs and, hence, does not require strong assumptions about the clutter and noise distributions. The proposed technique is evaluated using electromagnetic field simulation data of shallow-buried targets. We show that, compared to detectors based on parametric models, robust detectors can lead to significantly reduced false alarm rates, particularly in cases where there is a mismatch between the assumed model and the true distributions.
\end{abstract}


\IEEEpeerreviewmaketitle

\section{Introduction}
\label{sec:introduction}

\IEEEPARstart{L}{andmine} Monitor 2018 has reported over \text{\num{122000}} civilian casualties and injuries, caused by landmines and other unexploded ordnance (UXO), since it began global tracking in 1999 \cite{monitor}. It is inevitable that many more occurrences went unrecorded due to the lack of national monitoring systems in some affected countries. UXOs not only pose a threat to the lives and well-being of the population, but also hinder the economic development of a nation. Consequently, further improvements in techniques for reliable and safe detection of UXOs are of critical importance.

Ground-Penetrating Radar (GPR) is a nondestructive method that uses electromagnetic radiation to detect buried targets \cite{Daniels1996, Zoubir2002}. It can detect metallic as well as non-metallic targets having dielectric properties different from the background medium. A forward-looking GPR (FL-GPR) offers the potential of detecting landmines and UXOs with reduced risk to the operator \cite{Dogaru2012, Phelan2013, Comite2017, Comite2018}. This is accomplished with a large standoff distance between the detector and the targets, rendering it an attractive option compared to a conventional downward-looking radar that incurs the chance of disturbing the scene and activating the targets. However, a challenge of using FL-GPR is that the illuminating signals and the reflected signals experience substantial attenuation owing to the near cancellation of the direct and ground-reflected waves. Furthermore, the interface roughness and subsurface clutter, which are usually highly non-stationary, have a strong impact on FL-GPR performance. In order to reduce detection errors, these effects need to be compensated with an appropriate signal processing method.

Several methods to overcome the challenges of FL-GPR have been proposed \cite{Sun2003, Sun2005, Wang2007, Ton2010, Jin2012, Liao2014, Comite2017, Comite2018}. In \cite{Sun2003}, the authors consider the problem of detecting buried plastic landmines using FL-GPR and characterize the scattering phenomenon in the time-frequency domain. This motivated the use of a wavelet packet transform and a neural network classifier in \cite{Sun2005} to increase the detection performance. By dividing a full frequency band into several sub-bands and employing a multi-feature classifier, improved performance for detecting metallic landmines is achieved in  \cite{Wang2007}. In \cite{Ton2010}, an adaptive multi-transceiver imaging technique is proposed that effectively integrates imaging information obtained through platform motion. Feature extraction of the bistatic scattering images from a multiple-input-multiple-output radar system is proposed in \cite{Jin2012} to differentiate targets from clutter. A time-reversal imaging algorithm is proposed in \cite{Liao2014}, which employs three-dimensional finite-difference time-domain modeling to characterize the scattering from the ground surface and buried targets. The method proposed in \cite{Comite2017} forms a multiview tomographic image as a reference for performing a statistical parametric test to detect shallow-buried landmines in the presence of a rough ground surface. Furthermore, an adaptive approach is proposed in \cite{Comite2018} to reduce the false alarm rate by iteratively updating the estimated parameters of the clutter and target distributions under changing viewpoints. Another adaptive method for estimating the statistics of parameters with application to detecting objects in through-the-wall imaging can be found in \cite{Debes2010} \cite{Debes2009}.

In this paper, we propose a pixel-wise likelihood-ratio test (LRT) to detect landmines and UXOs in the image domain. In contrast to existing approaches, we design the test to be robust against statistical model deviations \cite{Zoubir2018}. More precisely, instead of modeling the distributions of the pixel intensity under each hypothesis with a parametric family of distributions, we use training data to construct two feasible bands within which the probability density functions (pdfs) under either hypothesis are assumed to lie. The detector is then designed such that it minimizes the maximum error probability for all possible density pairs within the two bands. In \cite{Kassam1981, Faus2016, Pambudi2018a, Pambudi2019}, it is shown that under very mild assumptions a minimax optimal test for such an uncertainty model is guaranteed to exist. The reason behind following a minimax approach is that accurate estimation of the distribution of the background clutter, given its non-stationary behavior, is highly challenging. The proposed technique overcomes this issue since it does not need an accurate estimate in the first place and is guaranteed to perform well over a set of feasible distributions. 

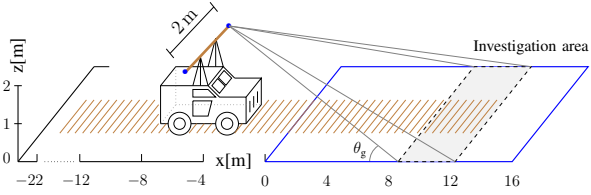
\begin{figure}[t]
	\centering
	\begin{minipage}[c]{0.9\linewidth}
		\resizebox{1\linewidth}{!} { 
\centering
\begin{tikzpicture}
\coordinate (zmax) at (0, 2);
\coordinate (xmax) at (14, 0);
\coordinate (y1) at (0,0);

\draw[thick] (-1,0)  -- (1,2.5);   
\draw [thick] (-1,0) -- (-1,2);    
\draw [thick] (-1,0) -- (-0.5,0);   
\draw [dotted] (-0.3,0) -- (0.5, 0); 
\draw [thick] (0.625,0) -- (3.875,0);   

\node [left of = y1, yshift = 0.1 cm, xshift = -0.3 cm]{\large $0$};
\node [left of = y1, yshift = 1 cm, xshift = -0.3  cm]{\large $1$};
\node [left of = y1, yshift = 2 cm, xshift = -0.3  cm]{\large $2$};

\draw ($(y1) + (-1.1,1)$) -- ++ (0.2,0);
\draw ($(y1) + (-1.1,2)$) -- ++ (0.2,0);
\node [left of = y1, yshift = 2.8 cm, xshift = 0 cm] {\rotatebox{90} {\Large z[m]}};
\draw ($(y1) + (1,2.5)$) -- ++ (0.4,0);

\draw ($(y1) + (-0.5,0)$) -- ++ (0,0.2);
\draw ($(y1) + (0.625,0)$) -- ++ (0,0.2);
\draw ($(y1) + (2.25,0)$) -- ++ (0,0.2);
\draw ($(y1) + (3.875,0)$) -- ++ (0,0.2);
\draw ($(y1) + (5.5,0)$) -- ++ (0,0.2);

\node [below of = y1, xshift = -0.7 cm, yshift = 0.5 cm] {\large $-22$};

\node [below of = y1, xshift = 0.525 cm, yshift = 0.5 cm] {\large $-12$};
\node [below of = y1, xshift = 2.15 cm, yshift = 0.5 cm] {\large $-8$};
\node [below of = y1, xshift = 3.675 cm, yshift = 0.5cm] {\large $-4$};

\node [below of = y1, xshift = 5.5 cm, yshift = 0.5cm] {\large $0$};
\node [below of = y1, xshift = 7.125 cm, yshift = 0.5cm] {\large $4$};
\node [below of = y1, xshift = 8.75 cm, yshift = 0.5cm]  {\large $8$};
\node [below of = y1, xshift = 10.375 cm, yshift = 0.5cm]  {\large $12$};
\node [below of = y1, xshift = 12 cm, yshift = 0.5cm]  {\large $16$};
\node [below of = y1, xshift = 4.675 cm, yshift = 1 cm]  {\Large x[m]};
\node at (8,0.3)  {\large $\theta_\text{g}$};
\draw [gray] (8.25,0) arc (180:125:0.5cm);

\draw[line width=0.3 mm, blue] (5.5,0) -- (12,0) -- (14,2.5) -- (7.5,2.5) -- (5.5,0);
\node [above of = y1, xshift = 12.5 cm, yshift = 2 cm]  {\large {Investigation area}};

\filldraw[fill=black!5!white, draw=black, dashed] (9,0) -- (10.5,0) -- (12.5,2.5) --(11,2.5) -- (9,0);

\draw[black] (5,1) -- (5,1.7) -- (4.3, 1.7) -- (4,2) -- (2.75,2) -- (2.75,1);

\draw[black] (2.75,1) -- (2.95,1);
\draw[black] (5,1) -- (4.8,1);
\draw[black] (3.55,1) -- (4.2,1);

\draw (3.25,1) circle (0.3cm);
\draw (3.25,1) circle (0.15cm);
\draw (4.5,1) circle (0.3cm);
\draw (4.5,1) circle (0.15cm);

\draw[black] (5,1) -- (5.4,1.5) -- (5.4,2.2) -- (5,1.7);

\draw[black] (2.75,2) -- (3.15,2.5) -- (4.4,2.5) -- (4,2);

\draw[black] (4.4,2.5) -- (4.7,2.2);

\draw[black] (4.7, 2.2) -- (5.4,2.2);

\draw[black] (3.8,2) -- (3.8,2.75);
\draw[black] (4,2) --  (3.8,2.75);
\draw[black] (3.6,2) -- (3.8,2.75);

\draw[black] (4.2,2.5) -- (4.2,3.25);
\draw[black] (4.4,2.5) -- (4.2,3.25);
\draw[black] (4,2.5) -- (4.2,3.25);

\draw [brown, line width=0.7 mm]  (3.4, 2.35) -- (4.55,3.573);
\node at (3.4, 2.35) {\color{blue}{$\bullet$}};
\node at (4.55,3.573) {\color{blue}{$\bullet$}};

\draw [black]    (3, 2.85) -- (4.15,4.073);
\draw [black]    (3.098, 2.73) -- (2.90,2.97);
\draw [black]    (4.052, 4.196) -- (4.25,3.95);
\node at (3.375, 3.7115)  {\rotatebox{47.34} {\Large {\SI{2}{\meter}}}};

\draw[black] (4.25,2.1875) -- (4.1,2) -- (4.3,1.8)-- (4.45,1.9875) -- (4.25,2.1875);
\draw[black] (4.42,2.4) -- (4.27,2.2125) -- (4.47,2.0125) -- (4.62,2.2) -- (4.42,2.4);

\draw[black] (5.1,1.3)  -- (5.1,1.6);
\draw[black] (5.3,1.55) -- (5.3,1.85);
\draw[black] (5.2,1.725) -- (5.2,1.425);

\draw[black] [dotted] [gray] (3.15,2.5)  -- (3.15,1.5) -- (2.75,1);
\draw[black] [dotted] [gray] (3.15,1.5)  -- (5.4,1.5);

\draw[black] (4,1.9)  -- (3.6,1.9) -- (3.6,1.7) -- (4.2,1.7) -- (4,1.9);
\draw[black] (3.6,1.6)  -- (3.6,1.2) -- (4,1.2) -- (4.1,1.6) -- (3.6,1.6);

\draw[gray] (4.55,3.573)  -- (10.5,0);
\draw[gray] (4.55,3.573)  -- (9,0);
\draw[gray] (4.55,3.573)  -- (12.5,2.5);
\draw[gray] (4.55,3.573)  -- (11,2.5);

\draw[brown] (0.1,0.75)  -- (0.8,1.625);
\draw[brown] (0.3,0.75)  -- (1,1.625);
\draw[brown] (0.5,0.75)  -- (1.2,1.625);
\draw[brown] (0.7,0.75)  -- (1.4,1.625);
\draw[brown] (0.9,0.75)  -- (1.6,1.625);
\draw[brown] (1.1,0.75)  -- (1.8,1.625);
\draw[brown] (1.3,0.75)  -- (2,1.625);
\draw[brown] (1.5,0.75)  -- (2.2,1.625);
\draw[brown] (1.7,0.75)  -- (2.4,1.625);
\draw[brown] (1.9,0.75)  -- (2.6,1.625);
\draw[brown] (2.1,0.75)  -- (2.7,1.5);
\draw[brown] (2.3,0.75)  -- (2.7,1.25);
\draw[brown] (2.5,0.75)  -- (2.7,1);
\draw[brown] (2.7,0.75)  -- (2.9,1);
\draw[brown] (2.9,0.75)  -- (3,0.875);
\draw[brown] (3.5,0.75)  -- (3.7,1);
\draw[brown] (3.7,0.75)  -- (3.9,1);
\draw[brown] (3.9,0.75)  -- (4.1,1);
\draw[brown] (4.1,0.75)  -- (4.2,0.875);
\draw[brown] (4.3,0.75)  -- (4.32,0.775);
\draw[brown] (4.7,0.75)  -- (4.9,1);
\draw[brown] (4.9,0.75)  -- (5.6,1.625);
\draw[brown] (5.1,0.75)  -- (5.8,1.625);
\draw[brown] (5.3,0.75)  -- (6,1.625);
\draw[brown] (5.5,0.75)  -- (6.2,1.625);
\draw[brown] (5.7,0.75)  -- (6.4,1.625);
\draw[brown] (5.9,0.75)  -- (6.6,1.625);
\draw[brown] (6.1,0.75)  -- (6.8,1.625);
\draw[brown] (6.3,0.75)  -- (7,1.625);
\draw[brown] (6.5,0.75)  -- (7.2,1.625);
\draw[brown] (6.7,0.75)  -- (7.4,1.625);
\draw[brown] (6.9,0.75)  -- (7.6,1.625);
\draw[brown] (7.1,0.75)  -- (7.8,1.625);
\draw[brown] (7.3,0.75)  -- (8,1.625);
\draw[brown] (7.5,0.75)  -- (8.2,1.625);
\draw[brown] (7.7,0.75)  -- (8.4,1.625);
\draw[brown] (7.9,0.75)  -- (8.6,1.625);
\draw[brown] (8.1,0.75)  -- (8.8,1.625);
\draw[brown] (8.3,0.75)  -- (9,1.625);
\draw[brown] (8.5,0.75)  -- (9.2,1.625);
\draw[brown] (8.7,0.75)  -- (9.4,1.625);
\draw[brown] (8.9,0.75)  -- (9.6,1.625);
\draw[brown] (9.1,0.75)  -- (9.8,1.625);
\draw[brown] (9.3,0.75)  -- (10,1.625);
\draw[brown] (9.5,0.75)  -- (10.2,1.625);
\draw[brown] (9.7,0.75)  -- (10.4,1.625);
\draw[brown] (9.9,0.75)  -- (10.6,1.625);
\draw[brown] (10.1,0.75)  -- (10.8,1.625);
\draw[brown] (10.3,0.75)  -- (11,1.625);
\draw[brown] (10.5,0.75)  -- (11.2,1.625);
\draw[brown] (10.7,0.75)  -- (11.4,1.625);
\draw[brown] (10.9,0.75)  -- (11.6,1.625);
\end{tikzpicture}}
	\end{minipage}
	\caption{FL-GPR vehicle-based system with antenna array mounted on top. Brown lines on the ground indicate sensor array positions.}
	\label{fig_FLGPR}
\end{figure}

The remainder of this paper is organized as follows. Section~\ref{sec:system_and_measurements} describes the configuration of the FL-GPR system and the measurement setup. Section~\ref{sec:problem_formulation} formulates the target detection problem as a statistical hypothesis test and discusses the differences between the parametric LRT previously studied in \cite{Comite2017, Comite2018, Debes2009} and the proposed minimax approach. It also addresses the question of how to construct the least favorable densities (LFD), which play a central role in designing the minimax robust detector. Section~\ref{sec:uncertainty_sets} first discusses how the training data is obtained and how it is used to estimate the target and clutter distributions in existing parametric approaches. Based on these results, two approaches for constructing the uncertainty sets of feasible distributions for the robust detector are detailed, both being special cases of the density band model. In Section~\ref{sec:simulation_results}, numerical results in different rough surface environments are presented and the performance of the proposed detector is compared to that of parametric alternatives. Finally, Section~\ref{sec:conclusion} concludes the paper. 

\section{FL-GPR System and Measurement Configuration}
\label{sec:system_and_measurements}

We work with numerical data simulated using the Near-Field Finite-Difference Time-Domain software package, NAFDTD, developed by the U.S.~Army Research Laboratory (ARL) \cite{Dogaru2012}. Fig.~\ref{fig_FLGPR} illustrates the FL-GPR system with a \SI{2}{\meter} wide antenna array, comprising 2 transmitters and 16 receivers, mounted on top of a vehicle at the approximate height of \SI{2}{\meter}. In order to sense the investigation area (\SI{16}{\meter} $\times$ \SI{10}{\meter} blue rectangle in Fig.~\ref{fig_FLGPR}), the radar platform moves along the x-direction, starting at \SI{-22}{\meter}.  The FL-GPR system operates in forward-looking mode from $x=\SI{-22}{\meter}$ to $x=\SI{11}{\meter}$, sensing the investigation area at grazing angle $\theta_\text{g} \in [\ang{5}, \ang{20}]$ approximately with a stepped frequency signal covering the  \SI{0.3}{} -- \SI{1.5}{\GHz} band in \SI{6}{\MHz} increments.

\begin{figure*}[ht]
	\centering
	\resizebox{0.8\linewidth}{!} {\input{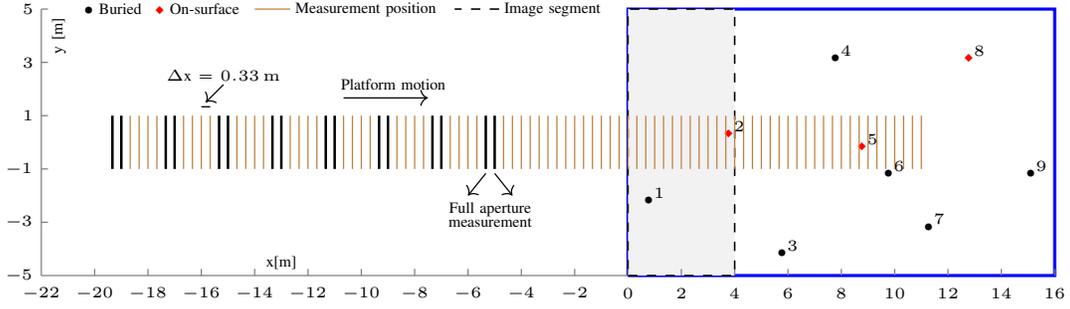}}
	\caption{Top view of the FL-GPR measurement. The blue rectangle indicates the investigation area containing nine targets, buried or placed on ground surface. Eight full aperture measurements are integrated to image the first segment (dashed line rectangle), corresponding to the farthest viewpoint.}
	\label{fig_measurement}
\end{figure*}

Fig.~\ref{fig_measurement} depicts the measurement configuration considered in the simulation. The investigation area is a rough surface environment containing nine targets.  The types of targets are detailed in Table~\ref{target_list}. Six targets are buried at a depth of 3 cm, five of which are metallic landmines \{1, 3, 4, 6, 7\} and one is made of plastic \{9\}. The remaining targets, two plastic landmines \{2 and 8\} and a metallic one \{5\}, are placed on the surface. The considered on-surface and shallow-buried target deployments mimic realistic scenarios, typical of anti-personnel landmines and/or anti-tank landmines laid by specialized minelayer vehicles \cite{ukarmy}. Plastic landmines are characterized with a relative dielectric constant $\varepsilon_r = 3.1$ and conductivity $\sigma = \SI{2}{\milli\siemens.\per\meter},$ which is representative of a minimum-metallic mine \cite{m14}. The ground is modeled as a dielectric medium which is non-dispersive, non-magnetic, and homogeneous with $\varepsilon_r = 6$ and $\sigma = \SI{10}{\milli\siemens.\per\meter}$. These electrical properties are typical of medium-dry soil \cite{itu}. The surface roughness is described as a two-dimensional Gaussian random process parameterized by the root mean square height $h_{rms}$ and the correlation length $l_c$ \cite{Comite2017}. Here, we consider two surface roughness profiles: ($h_{rms} = \SI{0.8}{\centi\meter}$, $l_c = \SI{14.26}{\centi\meter}$) and  ($h_{rms} = \SI{1.6}{\centi\meter}$, $l_c = \SI{14.93}{\centi\meter}$). 

\begin{table}[t]
	\renewcommand{\arraystretch}{1.3}
	\caption{\normalsize List of targets and their types.}
	\label{target_list}
	\centering
	\resizebox{0.9\columnwidth}{!}{%
	\begin{tabular}{|c| l || c | l |}
		\hline
		\cellcolor{g} \bfseries \small No. & \cellcolor{g} \bfseries \normalsize Mine type & \cellcolor{g} \bfseries \small No. & \cellcolor{g} \bfseries \normalsize Mine type  \\
		\hline\hline
		\small\{1\}		& \small Metallic anti-personnel  & \small\{4,5\}	& \small Metallic anti-tank \\
		\hline
		\small\{2,8\} 	& \small Plastic anti-personnel & \small\{9\} 	& \small Plastic anti-tank \\
		\hline
		\small\{3,6,7\}	& \small Metallic \SI{155}{}\SI{}{\milli\meter} shell & &\\
		\hline
	\end{tabular}
	}
\end{table}

\begin{figure}[t]
	\centering
	\begin{subfigure}[t]{0.8\linewidth}
		\begin{minipage}{1\linewidth}
			\resizebox{1\linewidth}{!} {\input{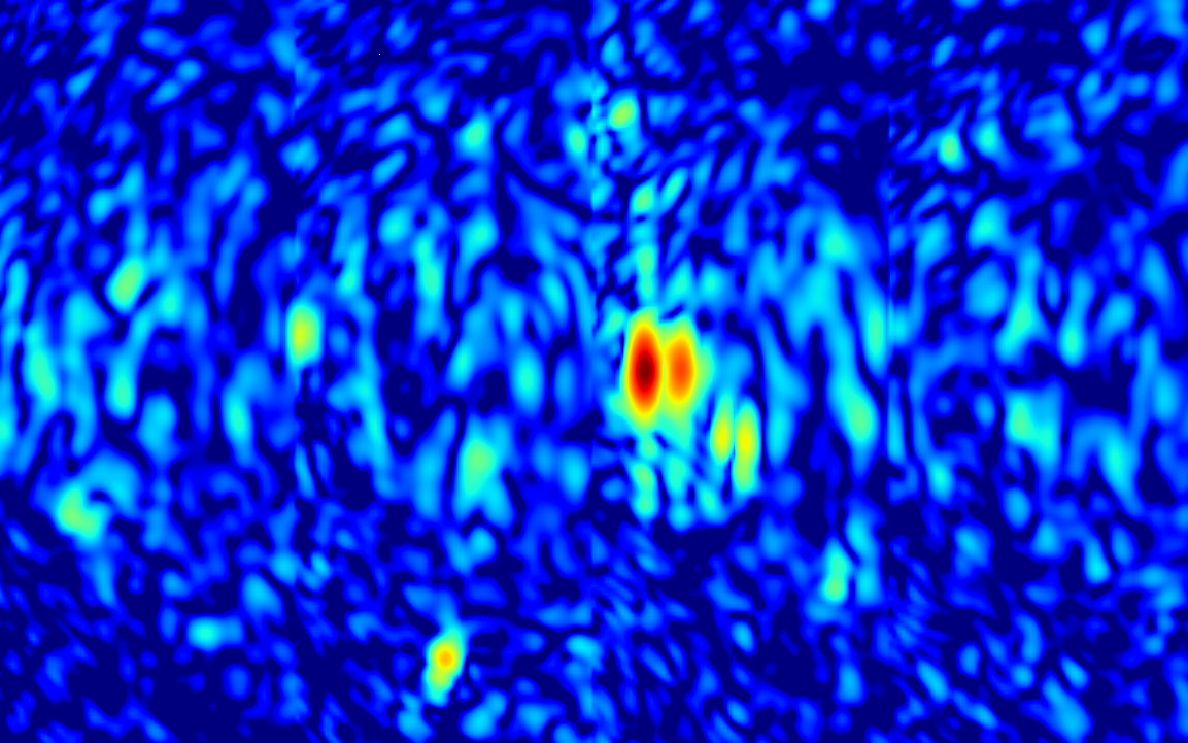}}
			\caption{Farthest viewpoint}
			\label{fig_tom1a}
		\end{minipage}
	\end{subfigure}
	\begin{subfigure}[t]{0.8\linewidth}
		\begin{minipage}{1\linewidth}
			\resizebox{1\linewidth}{!} {\input{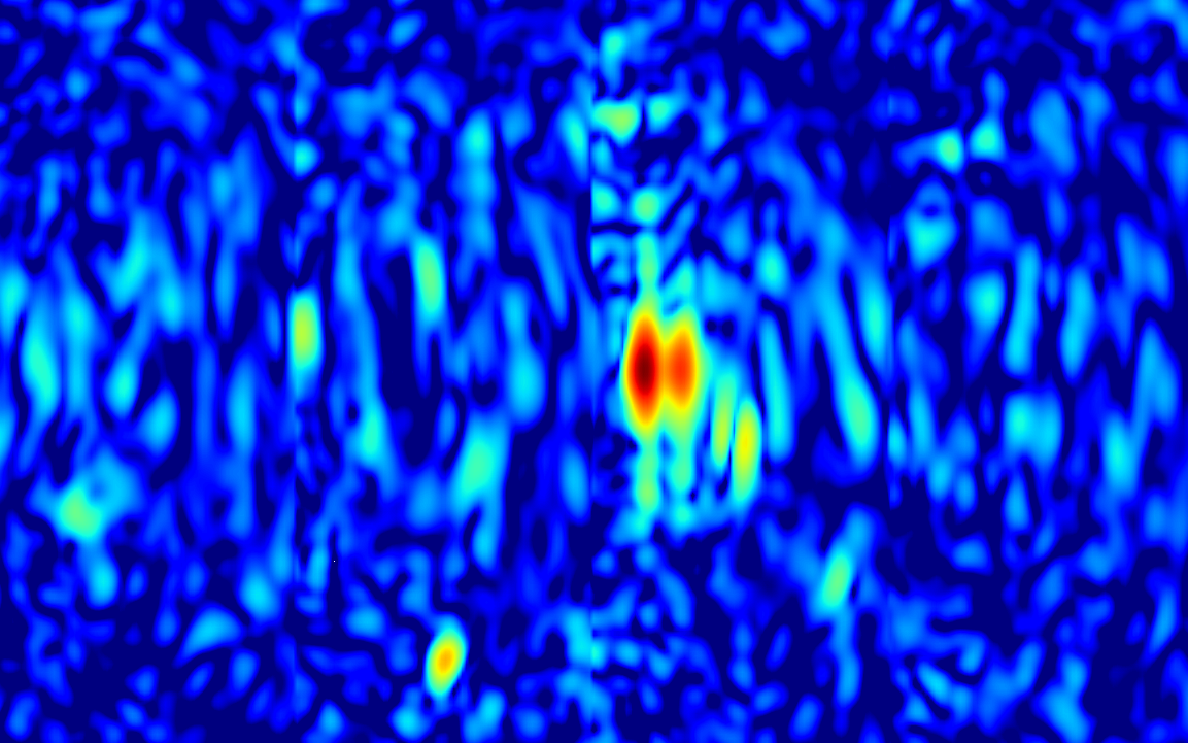}}
			\caption{Closest viewpoint}
			\label{fig_tom2a}
		\end{minipage}
	\end{subfigure}
	\caption{Tomographic images for low surface roughness profile ($h_\text{rms}=$ \SI{0.8}{\centi\meter}, $l_c=$ \SI{14.26}{\centi\meter}). White crosses indicate targets.}
	\label{fig_toma}
\end{figure}
\begin{figure}[t]
	\centering
	\begin{subfigure}[]{0.8\linewidth}
		\begin{minipage}{1\linewidth}
			\resizebox{1\linewidth}{!} {\input{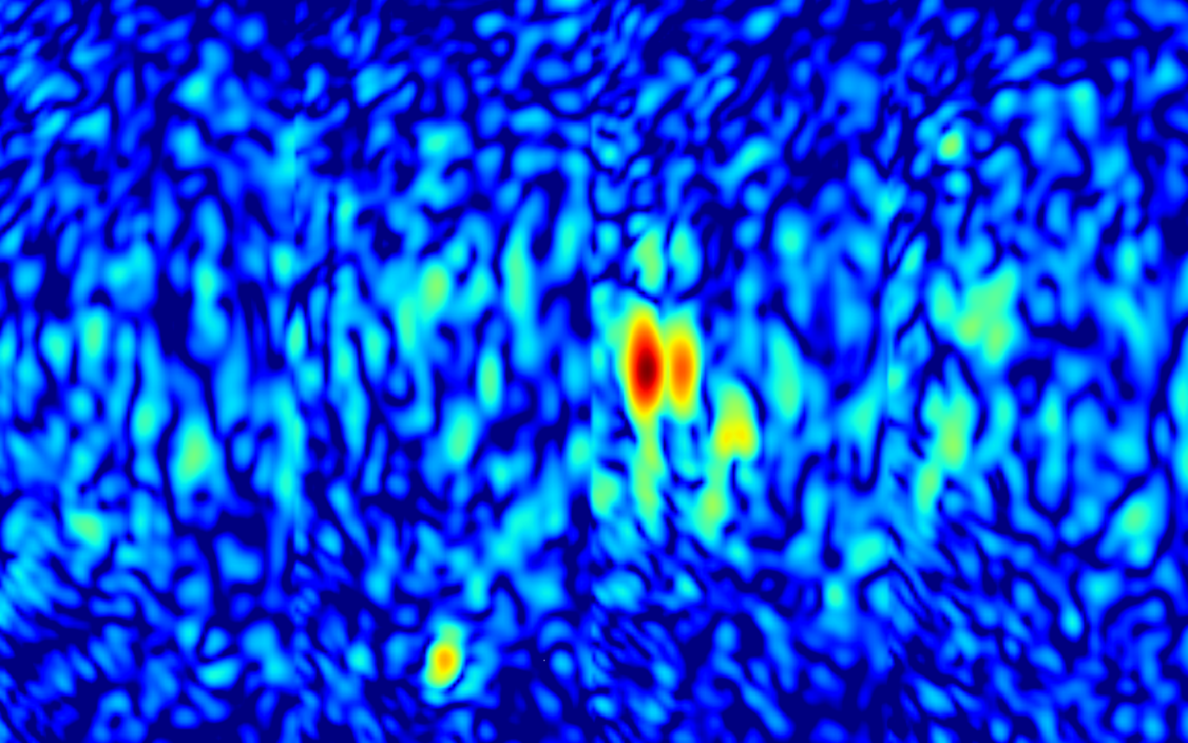}}
			\caption{Farthest viewpoint}
			\label{fig_tom1b}
		\end{minipage}
	\end{subfigure}
	\begin{subfigure}[]{0.8\linewidth}
		\begin{minipage}{1\linewidth}
			\resizebox{1\linewidth}{!} {\input{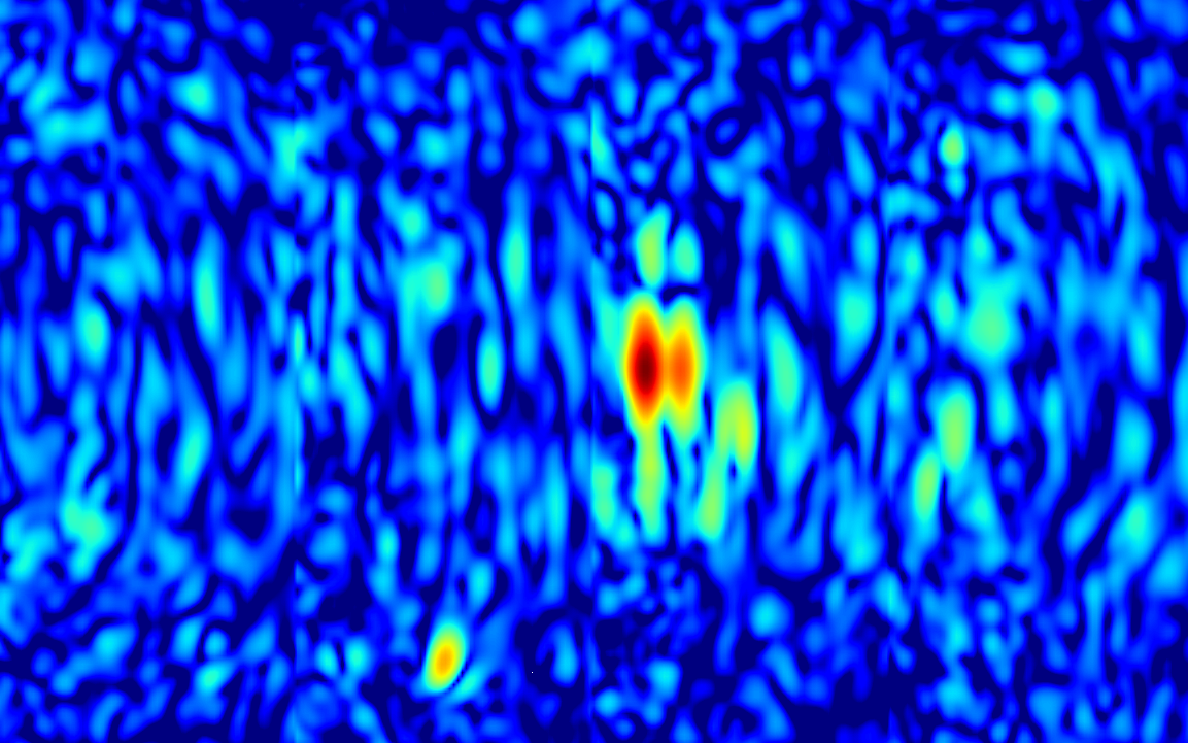}}
			\caption{Closest viewpoint}
			\label{fig_tom2b}
		\end{minipage}
	\end{subfigure}
	\caption{Tomographic images for high surface roughness profile ($h_\text{rms}=$ \SI{1.6}{\centi\meter}, $l_c=$ \SI{14.93}{\centi\meter}).}
	\label{fig_tomb}
\end{figure}

A total of 90 array positions spaced $\Delta$x = \SI{0.33}{\meter} apart are considered, whose projections on the x-y plane are depicted as parallel brown lines in Fig.~\ref{fig_measurement}. Two adjacent array positions realize a full aperture measurement. In each array position, we activate only one transmit element, while all receive elements simultaneously record the reflected signals from the scene. A tomographic image is constructed by coherently integrating multiple full aperture measurements. The tomographic image is composed of four image segments, each of dimension \SI{4}{\meter} $\times$ \SI{10}{\meter}. By integrating eight full aperture measurements with each pair separated by four array positions, we obtain an image segment. Fig.~\ref{fig_measurement} indicates the eight full aperture measurements between \SI{-19.33}{} and \SI{-5}{\meter} used to image the first segment (dashed rectangle in Fig.~\ref{fig_measurement}). The full aperture measurements used to construct the second, third, and fourth image segments range from \SI{-15.33}{} to \SI{-1}{\meter}, \SI{-11.33}{} to \SI{3}{\meter}, and \SI{-7.33}{} to \SI{7}{\meter}, respectively. However, these are not depicted in Fig.~\ref{fig_measurement}. A tomographic image is thus constructed by integrating \num{32} full aperture measurements from $\text{x} =\SI{-19.33}{\meter}$ to $\text{x} =\SI{7}{\meter}$. In order to improve the detection performance, we generate ten additional images by successively moving the radar closer to the target area. Compared to the image in Fig.~\ref{fig_toma} corresponding to the farthest viewpoint, the tomographic image from the viewpoint closest to the scene is realized by integrating measurements from $\text{x} =\SI{-16}{\meter}$ to $\text{x} = \SI{10.33}{\meter}$. By this means, we have a total image of $M=11$ providing various views of the investigation area. For more details of the multiview tomographic imaging approach, refer to \cite{Comite2017}.

The images in Figs.~\ref{fig_toma}~and~\ref{fig_tomb} are normalized to \SI{40}{\dB} dynamic range and each consists of $N_{\text{x}}=1153$ pixels in downrange and $N_{\text{y}}=721$ pixels in crossrange with a resolution of \SI{5}{\centi\meter}. The targets located near the boundaries of the investigation area are less visible since they are outside of the mainlobe of the antenna array whose physical aperture is smaller than the image extent in cross range. Compared to the landmines placed on the surface, the buried landmines are more challenging to discern due to the clutter caused by the radar back-scatter from the rough ground surface. In general, recognizing a plastic landmine is harder than a metallic one, since its return signal energy is comparable to that of clutter. The difference in clutter returns across the different viewpoints is clearly visible and illustrates the non-stationary behavior of the scattering from the rough ground surface. As expected, the higher surface roughness leads to stronger clutter echoes, thus making the detection more challenging. 

\section{Problem Formulation and The Likelihood-Ratio Test}
\label{sec:problem_formulation}

A tomographic image is represented as a two-dimensional array of normalized pixel intensities
\begin{equation*}
  x(i,j), \quad 1 \leq i \leq N_{\text{x}}, \enspace 1\leq j \leq N_{\text{y}},
\end{equation*}
where $x(i,j) \in [0,1]$. The detection problem is to decide whether a pixel at $(i,j)$ belongs to target or clutter. The problem can be defined as a test between a null and an alternative hypothesis as,
\begin{equation}
  \begin{aligned} 
    H_0 : \enspace& \text{no target present},\\
    H_1 : \enspace& \text{target present}.
  \end{aligned}
  \label{hypo}
\end{equation}
The proposed robust test, as well as the existing parametric LRT \cite{Comite2017, Comite2018}, which is used as a reference for comparison, are based on the Neyman--Pearson approach \cite{Lehmann2005} to optimal test design. That is, the test is designed such that it minimizes the miss detection (type II error) probability under a constraint on the false alarm (type I error) probability. In other words, it minimizes the probability of confusing a target pixel for a clutter pixel for a given probability of confusing a clutter pixel for a target pixel. In order to highlight the similarities and differences between the parametric and the robust LRT, we briefly summarize the former before detailing the latter.

\subsection{Parametric Likelihood-Ratio Test}

For simplicity, let us assume that the pixel intensities of all $M$ tomographic images are independently and identically distributed (iid). For the $m$th tomographic image, a pixel-wise LRT is given by
\begin{equation}
  L_m(i,j) = \frac{p_1(x_m(i,j))}{p_0(x_m(i,j))} \; \mathop{\gtrless}_{H_0}^{H_1} \; \gamma, 
  \label{eq_lrt}
\end{equation}
where $x_m(i,j)$ denotes the intensity of the pixel at coordinate $(i,j)$ of the $m$th image and ${p_s}$ denote the conditional probability density functions of the pixel intensities under hypotheses $H_s$, $s= 0,\, 1$. The threshold $\gamma$ is determined based on the targeted false alarm probability $\alpha$ and can be determined by solving 
\begin{equation}
  \alpha = \intop_{\gamma}^{\infty }f_L(l|H_0) \, \mathrm{d}l
  \label{eq_fa}
\end{equation}
for $\gamma$. Here, $f_L(l|\, H_0)$ denotes the density of the likelihood ratio $L$ under the null hypothesis. In \cite{Comite2017, Comite2018}, the authors show that the pixel intensities approximately follow a Rayleigh distribution under $H_0$ and a Gaussian mixture distribution under $H_1$. The corresponding density functions are given by
\begin{equation}
  \begin{aligned}
    p_0(x) &= x \, / \sigma_0^{2} \, \cdot \exp \, (-x \, / 2\sigma_0^{2}), \\
    p_1(x) &= \sum_{k=1}^{K} \phi_{k} \, \mathcal{N}(x \: | \: \mu_{k}, \sigma _k^2), 
  \end{aligned}
  \label{eq_nom}
\end{equation}
where $\sigma_0^{2}$ is the parameter of the Rayleigh distribution, $\mathcal{N}(\bullet \: | \: \mu, \sigma^2)$ denotes a Gaussian distribution with mean $\mu$ and variance $\sigma^2$, $K$ is the number of Gaussian distributions in the mixture, and $\phi_{1}, \ldots, \phi_K$ are mixture weights. Based on this model, the authors in \cite{Comite2017, Comite2018, Seng2012} propose a generalized LRT, i.e., the test statistic in \eqref{eq_lrt} is evaluated by replacing all unknown parameters in \eqref{eq_nom} with their maximum likelihood estimates.

\subsection{Minimax Robust Likelihood-Ratio Test}

The idea of minimax robust hypothesis testing is to design a test such that it works well under all feasible distributions. That is, in contrast to the generalized LRT, which tries to \emph{reduce} the uncertainty by estimating unknown parameters, a minimax robust test \emph{tolerates} uncertainty.

We start by considering the general case of two composite hypotheses 
\begin{equation}
  \begin{aligned} 
    H_0 : \enspace& P \in \mathscr{F}_0,\\
    H_1 : \enspace& P \in \mathscr{F}_1,
  \end{aligned}
  \label{eq_hypo_robust}
\end{equation}
where $\mathscr{F}_0$ and $\mathscr{F}_1$ are two disjoint sets of feasible distributions which are chosen such that they adequately capture the distributional uncertainties of the underlying detection problem. Here, we use Kassam's band model for this purpose, which is a generalization of the $\varepsilon$-contamination model \cite{Huber1965} and has been shown to provide a good trade-off between flexibility and tractability \cite{Kassam1981, Faus2016}. It is given by 
\begin{equation}
  \begin{aligned}
    \mathscr{F}_0 = \{ p_0\,\,|\,\, p_{0}^{\prime}(x) \leq p_0(x) \leq p_{0}^{\prime \prime}(x)\, \},\\
    \mathscr{F}_1 = \{ p_1\,\,|\,\, p_{1}^{\prime}(x) \leq p_1(x) \leq p_{1}^{\prime \prime}(x)\, \},
  \end{aligned}
  \label{eq_band}
\end{equation}
where $p_{s}^{\prime}$ and $p_{s}^{\prime \prime}$ denote lower and upper bounds on the true density, respectively. For the problem at hand, the bounds need to be non-negative functions satisfying 
\begin{equation}\label{key}
  \int_{}^{} p_{s}^{\prime}(x)\, \mathrm{d}x \leq 1 \leq \int_{}^{} p_{s}^{\prime \prime}(x)\, \mathrm{d}x, \quad s = 0,1,
\end{equation}
but can otherwise be chosen freely by the test designer.

In principle, a minimax robust test is designed by finding a pair of densities $(g_0, g_1) \in \mathscr{F}_0 \times \mathscr{F}_1$ which is \emph{least favorable} in the sense that it simultaneously maximizes both error probabilities among all feasible densities. If such a pair exists, the corresponding minimax optimal test can be shown to be a threshold test whose test statistic is the likelihood ratio of the least favorable densities (LFDs) \cite{Huber1965, Faus2016}. In \cite{Faus2016}, it is shown that the LFDs for the band model can, in general, be written as
\begin{equation}
  \begin{aligned}
    g_0(x)& = \min \: \lbrace p^{\prime \prime }_0(x) \,,\, \max \: \lbrace a_0 \, g_1(x), \, p^{\prime}_0(x) \rbrace \rbrace,\\ 
    g_1(x)& = \min \: \lbrace p^{\prime \prime }_1(x) \,,\, \max \: \lbrace a_1 \, g_0(x), \, p^{\prime}_1(x) \rbrace \rbrace,
  \end{aligned}
  \label{eq:lfd1}
\end{equation} 
where the two constants $a_0$ and $a_1$ have to be calculated such that the LFDs are valid densities, i.e., they integrate to one. Based on the procedure outlined in \cite{Faus2016}, an iterative algorithm to construct the LFDs is provided in Table~\ref{algorithm}. Starting with an initial guess for the LFDs ($g_0^0$, $g_1^0$), the algorithm alternately updates $g_0^n$ and $g_1^n$ by finding a root of
\begin{equation}
  \begin{aligned}
    f_s(a; g) =& \int \min \lbrace p_s^{\prime\prime}(x), \, \max \lbrace a\, g(x),\, p_s^{\prime}(x) \rbrace \rbrace \, dx - 1,
  \end{aligned}
  \label{density_criterion}
\end{equation}
which, for a given density $g$, is a non-decreasing function of the scalar $a$. The iteration is terminated once both densities have converged within a tolerance $\delta$.

\begin{table}[t]
  \renewcommand{\arraystretch}{1.3}
  \caption{Algorithm to construct the LFDs}
  \label{algorithm}
  \centering
  \resizebox{0.75\columnwidth}{!}{%
  \begin{tabular}{|l|}
    \hline
    \textbf{Input:} $p_0^{\prime},\, p_0^{\prime\prime},\, p_1^{\prime},\, p_1^{\prime\prime},\, g_0^0,\, g_1^0,\, \delta$\\
    \hline
    \textbf{Output:} $g_0$, $g_1$\\
    \hline
    \enspace 1: \textbf{repeat}\\
    \enspace 2: \qquad $a_0  \,\,\,\,\,\, = \text{root of  }f_0(a; g_1^n)$\\
    \enspace 3: \qquad $g_0^{n+1} = \min \: \lbrace p^{\prime \prime }_0\,,\, \max \: \lbrace a_0 \, g_0^{n}, \, p^{\prime }_0\rbrace \rbrace$\\
    \enspace 4: \qquad $a_1  \,\,\,\,\,\, = \text{root of  }f_1(a; g_0^{n+1})$\\
    \enspace 5: \qquad $g_1^{n+1} = \min \: \lbrace p^{\prime \prime }_1\,,\, \max \: \lbrace a_1 \, g_0^{n+1}, \, p^{\prime }_1\rbrace \rbrace$\\
    \enspace 6: \qquad 	\textbf{if} $\lVert g_0^{n} - g_0^{n+1} \rVert < \delta$ and $\lVert g_1^{n} - g_1^{n+1} \rVert < \delta$\\
    \enspace 7: \qquad \qquad \textbf{return} $g_0^{n}$, $g_1^{n}$ \\
    \enspace 8: \qquad $g_0^{n}$ = $g_0^{n+1}$\\
    \enspace 9: \qquad $g_1^{n}$ = $g_1^{n+1}$\\
    \hline
  \end{tabular}
  }
\end{table}

Note that the likelihood ratio of the least favorable densities, $g_1(x)/g_0(x)$, can take six possible values, namely
\begin{equation}
\frac{g_1(x)}{g_0(x)} \in \bigg\lbrace \, \frac{p^{\prime}_1(x)}{ p^{\prime}_0(x)},\, \frac{p^{\prime}_1(x)}{ p^{\prime\prime}_0(x)},\, \frac{p^{\prime\prime}_1(x)}{ p^{\prime}_0(x)},\, \frac{p^{\prime\prime}_1(x)}{ p^{\prime\prime}_0(x)},\, a_1,\, \frac{1}{a_0} \, \bigg\rbrace .
\label{eq:lrt1}
\end{equation}
The first four values correspond to regions where $g_0$ or $g_1$ coincide with either their lower or their upper bound. Hence, for the two remaining cases it holds that either $p^{\prime }_1(x) < g_1(x) < p^{\prime\prime}_1(x)$, which by \eqref{eq:lfd1} implies that
\begin{equation}
  g_1(x) = a_1 \, g_0(x) \quad \Rightarrow \quad  \frac{g_1(x)}{g_0(x)} = a_1, 
  \label{eq:lrt2}
\end{equation}
or that $p^{\prime }_0(x) < g_0(x) < p^{\prime\prime}_0(x)$, which implies
\begin{equation}
  g_0(x) = a_0 \, g_1(x) \quad \Rightarrow \frac{g_1(x)}{g_0(x)} = \frac{1}{a_0}.
\label{eq:lrt3}
\end{equation}

We can obtain an interesting special case of the band model by letting $p^{\prime\prime}_0, p^{\prime\prime}_1 \rightarrow \infty$, so that the LFDs in (\ref{eq:lfd1}) reduce to 
\begin{equation}
  \begin{aligned}
    g_0(x) & = \max \: \lbrace a_0 \, g_1(x), \, p^{\prime }_0(x) \rbrace,\\ 
    g_1(x) & = \max \: \lbrace a_1 \, g_0(x), \, p^{\prime }_1(x) \rbrace,
  \end{aligned}
  \label{eq:lfd2}
\end{equation}
and the possible values of the likelihood ratio become
\begin{equation}
\frac{g_1(x)}{g_0(x)} \in \bigg\lbrace \, \frac{p^{\prime}_1(x)}{ p^{\prime}_0(x)},\, a_1,\, \frac{1}{a_0} \, \bigg\rbrace. 
\label{eq:lrt4}
\end{equation}
That is, the density band model reduces to the $\varepsilon$-contamination model, and the corresponding minimax optimal test becomes Huber's clipped likelihood-ratio test, which is robust against outliers \cite{Huber1965}.

\section{Construction of the Uncertainty Sets}
\label{sec:uncertainty_sets}

Having calculated the LFDs, the robust LRT is then of the form \eqref{eq_lrt}, with the parametric densities ($p_0, p_1$) replaced by the LFDs ($g_0, g_1$). First, however, the uncertainty sets $\mathscr{F}_0$ and $\mathscr{F}_1$ need to be constructed. In order to do so, we propose a hybrid approach that combines parametric density estimates constructed from training data with classic nonparametric uncertainty models, which is detailed in this section.

\subsection{Training Data}
\label{sec:image_properties}

\begin{figure}[t]
	\centering
  \resizebox{0.8\linewidth}{!} {\input{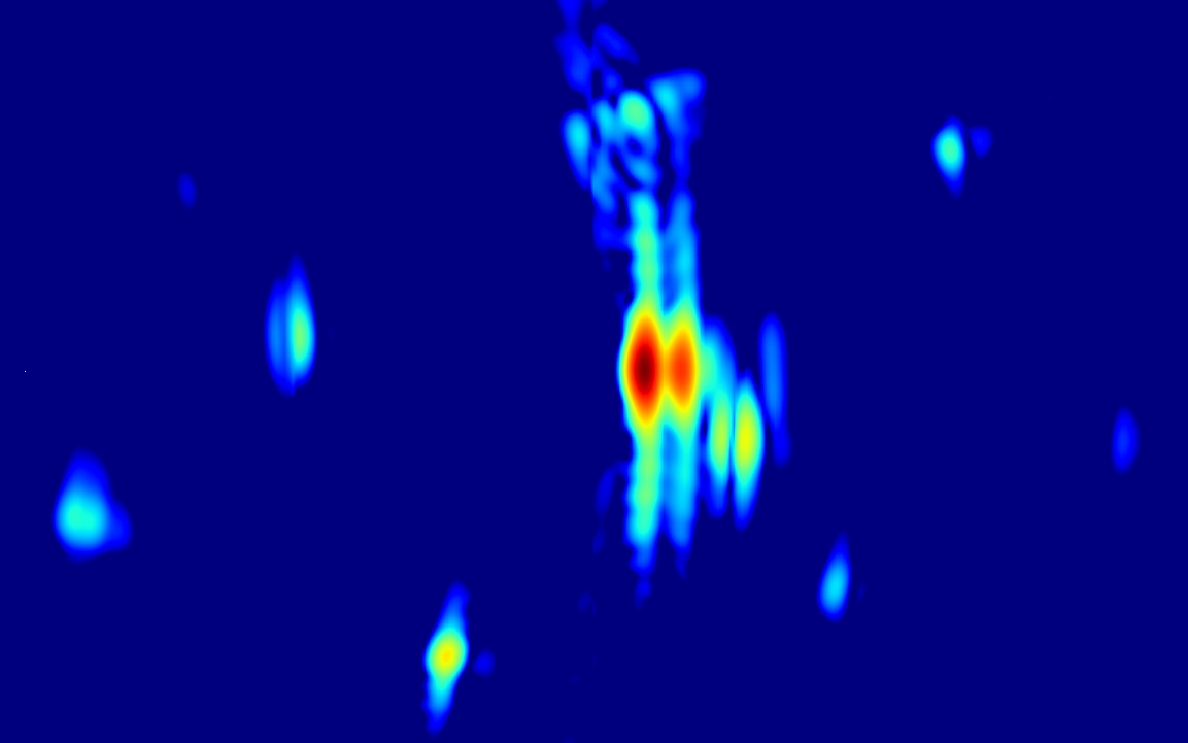}}
  \caption{Training image $T$. Normalized tomographic image of the closest viewpoint on flat ground surface.}
  \label{fig_freeclutter}
\end{figure}

In order to perform the LRT, \textit{a priori} knowledge about the distribution of the pixel intensities under each hypothesis is needed. In order to obtain this knowledge, we use two training images, $T$ and $C$, generated by the NAFDTD software. The image $T$ is a clutter-free image as shown in Fig.~\ref{fig_freeclutter}, which is a normalized tomographic image that was generated with the same simulation parameters as described in Section II, except that the ground was assumed to be flat. We then employ a region growing algorithm \cite{Adams1994, Rueda2007} to locate the target pixel regions. First, the center points of the nine target positions are chosen manually. The algorithm then successively adds adjacent pixels to the target region based on the connection criterion $\varphi \pm \text{\SI{3}{\dB}}$, where $\varphi$ denotes the pixel intensity of the seed points; see \cite{Adams1994, Rueda2007} for more details. The output of the region growing algorithm is a binary mask given by
\begin{equation}
  \mathrm{BM}(i,j) = \begin{cases}  
          1,  & \text{target region at}\, (i,j),\\
          0,  & \text{non-target region at}\, (i,j),
        \end{cases}
  \label{nask}
\end{equation}
with $i=1,..., N_{\text{x}}$ and $j=1,..., N_{\text{y}}$. By multiplying the original image $T$ with the mask $\mathrm{BM}$, we obtain the image $\textit{\textbf{T}}=T \cdot \mathrm{BM}$ that only contains the target pixels and zeros. 

The training image $C$ is the closest viewpoint tomographic image for the low surface roughness profile ($h_\text{rms}= \SI{0.8}{\centi\meter}$, $l_c = \SI{14.26}{\centi\meter}$) as shown in Fig.~\ref{fig_tom2a}. The image $C$ is multiplied by the mask image complement $\mathrm{BM}^c$ to obtain the image $\textit{\textbf{C}}$, which is composed of only clutter pixels and zeros. In this way, we obtain two data sets, $\mathcal{X}_t$ and $\mathcal{X}_c$, containing only target and clutter pixels, respectively:
\begin{eqnarray}
\begin{aligned}
\mathcal{X}_t & = \lbrace \, x\, |\, x\in  \textit{\textbf{T}}, \, x > 0 \, \rbrace,\\ 
\mathcal{X}_c & = \lbrace \, x\, |\, x\in  \textit{\textbf{C}}, \, x > 0 \, \rbrace.
\end{aligned}
\label{training pixel}
\end{eqnarray}
The data set $\mathcal{X}_t$ consists of $N_t = 3206$ pixels, while the total pixels in set $\mathcal{X}_c$ equals $N_c = N_{\text{xy}} - N_t,$ where $N_{\text{xy}}= N_{\text{x}} \times N_{\text{y}}$. In the next section, we detail how the uncertainty set needed for the robust likelihood-ratio test is constructed from these training data sets.

\begin{figure}[t]
	\centering
	\pgfsetplotmarksize{0pt}
	\resizebox{0.8\linewidth}{!} {%
	\begin{tikzpicture}
	  \tikzset{font={\fontsize{8.5pt}{12}\selectfont}}
	  \begin{axis}[ylabel style={at={(axis description cs: 0.07, 0.5)},anchor=south}, grid=major, xtick={0, 0.1, 0.2, 0.3, 0.4, 0.5, 0.6, 0.7, 0.8, 0.9, 1.0}, ytick={0, 5, 10, 15, 20, 25, 30, 35}, xlabel=$x$: pixel intensity, ylabel=Density, width=8.8 cm, height=5.5cm, legend entries={Clutter $\mathcal{X}_c$, Target $\mathcal{X}_t$}, legend style={at={(0.97,0.95)},anchor=north east},]
	  
	    \addplot +[brown!70, fill=brown!70, fill opacity=0.4, hist={density, bins=100, data min=0, data max=1}] table [y index=0,col sep=comma] {clutter_boot.csv};
	    \addplot +[blue!50, fill=blue!50, fill opacity=0.4, hist={density, bins=100, data min=0, data max=1}] table [y index=0] {target_boot.csv};
	    
	  \end{axis}
	\end{tikzpicture}
	}
	\caption{Histogram of the training data sets $\mathcal{X}_c$ and $\mathcal{X}_t$.}
	\label{histogram}
\end{figure}
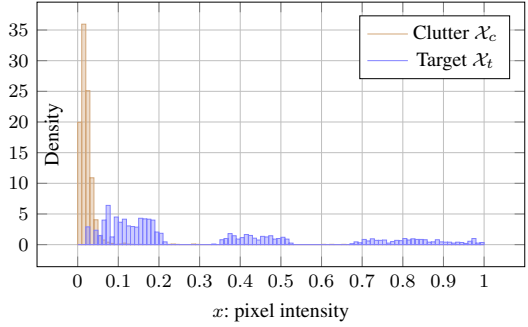

The histograms of the training data sets $\mathcal{X}_c$ and $\mathcal{X}_t$ are shown in Fig.~\ref{histogram}. As mentioned in Section~\ref{sec:problem_formulation}, the authors of \cite{Comite2017, Debes2009, Liao2012} showed that the clutter pixels are approximately Rayleigh distributed, while the distributions of the target pixels can be approximated by a Gaussian mixture distribution. Our aim is to design the robust test such that, on one hand, prior knowledge of the shape of the distributions can still be incorporated, but on the other hand, it is insensitive to deviations from the assumed model. This is accomplished by using the parametric distributions proposed in \cite{Comite2017, Debes2009, Liao2012} as \emph{nominal distributions} around which a neighborhood of feasible distributions is constructed. The robust detector is then designed such that it is minimax optimal over the set of feasible distributions. Two ways of choosing the neighborhood are detailed below: the density band model and the $\varepsilon$-contamination model.

\subsection{Density Band Model}

For the first uncertainty model, we construct a band of feasible densities as defined in \eqref{eq_band}. More precisely, the lower and upper bounds are chosen as
\begin{equation}
  p_s'(x) = 0.8\, p_s(x), \quad  p_s''(x) = 2.5\, p_s(x), \quad s = 0,1,
  \label{bounds1}
\end{equation}
where $p_0$ and $p_1$ denote the parametric density estimates in \eqref{eq_nom}. That is, we assume that the true distributions are close to the nominal distributions in the sense that the probability mass on any measurable set $E \subset \mathbb{R}$ is lower bounded by $0.8\, P_s(E)$ and upper bounded by $2.5\, P_s(E)$. Note that this implies that the deviations from the nominal model are continuous, i.e., events that are highly (un)likely under the nominal distribution are also highly (un)likely under all feasible distributions. The particular values of 0.8 and 2.5 were chosen heuristically based on a series of experiments. More specifically, they have been shown to correspond to a good compromise between guarding against slight but systematic model mismatches and guarding against severe but rare outliers \cite{Faus2016, Pambudi2018a}. 

Using the training data detailed above, we obtained maximum likelihood estimates for the parameters of the densities in \eqref{eq_nom} as shown in Table~\ref{parameter_estimates}. The corresponding bands of feasible densities are shown in Fig.~\ref{fig_band}, indicated by the shaded areas.
\begin{table}[t]
	\renewcommand{\arraystretch}{1.3}
	\caption{The parameter estimates for $p_0$, $p_1$.}
	\label{parameter_estimates}
	\centering
	\begin{tabular}{|c|c|c|c|c|}
		\hline
		\cellcolor{g} \boldmath $\hat{\sigma}_{0}$ & \cellcolor{g} \boldmath $\hat{\sigma}_{1}$ & \cellcolor{g} \boldmath $\hat{\sigma}_{2}$ & \cellcolor{g} \boldmath $\hat{\sigma}_{3}$ & \cellcolor{g} \boldmath $\hat{\phi}_{1}$ \\
		\hline
		0.025 & 0.050 & 0.048 & 0.088 &  0.601  \\
		\hline
		\hline
		\cellcolor{g} \boldmath $\hat{\phi}_{2}$ & \cellcolor{g} \boldmath $\hat{\phi}_{3}$  & \cellcolor{g} \boldmath $\hat{\mu}_{1}$  &  \cellcolor{g} \boldmath $\hat{\mu}_{2}$ & \cellcolor{g} \boldmath $\hat{\mu}_{3}$\\
		\hline
		0.212 & 0.186  & 0.117 & 0.430	& 0.833 \\
		\hline
	\end{tabular}
\end{table} 
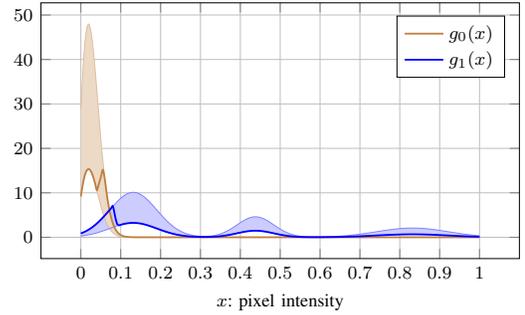
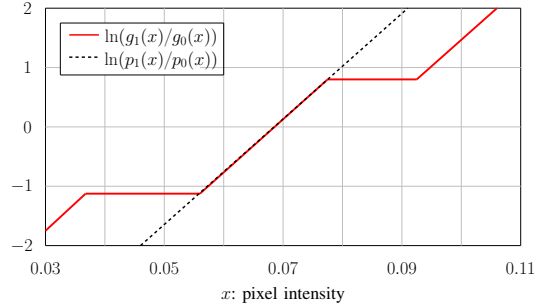
\begin{figure}[t]
	\centering
	\begin{subfigure}{0.9\linewidth}
		\centering
		\resizebox{0.87\linewidth}{!} {%
		\begin{tikzpicture}
		\tikzset{font={\fontsize{8pt}{11}\selectfont}}
		\begin{axis} [thin, no markers, width=9 cm, height=5.55cm, smooth, ylabel style={at={(axis description cs: 0.04,0.5)},anchor=south}, xtick={0, 0.1, 0.2, 0.3, 0.4, 0.5, 0.6, 0.7, 0.8, 0.9, 1.0},  xlabel={$x$: pixel intensity}, legend cell align={right}, grid=major, legend style={at={(0.97,0.95)},anchor=north east}, ytick={0, 10, 20, 30, 40, 50}, legend entries={, , , , $g_0(x)$, $g_1(x)$}]
		\addplot+[name path=A,brown!50] table [x=a, y=b, col sep=space] {data1.csv};
		\addplot+[name path=B,brown!50] table [x=a, y=c, col sep=space] {data1.csv};
		\addplot+[name path=C,blue!50] table [x=a, y=d, col sep=space] {data1.csv};
		\addplot+[name path=D,blue!50] table [x=a, y=e, col sep=space] {data1.csv};
		
		\addplot[color=brown] [thick] table [x=a, y=qo, col sep=space] {data1.csv};
		\addplot[color=blue]  [thick] table [x=a, y=q1, col sep=space] {data1.csv};
		
		\addplot[blue!20] fill between[of=C and D];
		\addplot[brown!30] fill between[of=A and B];
		\end{axis}
		\end{tikzpicture}
		}
		\caption{Density band model and LFDs}
		\label{fig_band}
	\end{subfigure}
	\begin{subfigure}{0.9\linewidth}
		\centering
		\resizebox{0.893\linewidth}{!} {\begin{tikzpicture}
\coordinate (ymax) at (0,8);
\coordinate (xmax) at (16, 0);
\coordinate (1) at (0,0);

\draw [thick] (0,0)  -- (ymax);
\draw [thick] (0,0)  -- (xmax);
\draw [thick] (16,8) -- (xmax);
\draw [thick] (16,8) -- (ymax);

\draw [thick] [lightgray](2,0) -- (2,5.75);
\draw [thick] [lightgray](2,7.5) -- (2,8);
\draw [thick] [lightgray](4,0) -- (4,5.75);
\draw [thick] [lightgray](4,7.5) -- (4,8);
\draw [thick] [lightgray](6,0) -- (6,8);
\draw [thick] [lightgray](8,0) -- (8,8);
\draw [thick] [lightgray](10,0) -- (10,8);
\draw [thick] [lightgray](12,0) -- (12,8);
\draw [thick] [lightgray](14,0) -- (14,8);

\draw [thick] [lightgray](0,2) -- (16,2);
\draw [thick] [lightgray](0,4) -- (16,4);
\draw [thick] [lightgray](5.2,6) -- (16,6);
\draw [thick] [lightgray](0,6) -- (0.5,6);

\draw [line width=0.5mm][red] (0.8,7.1)    -- (1.8,7.1);
\draw [line width=0.5mm] [dashed][black]  (0.8,6.35) -- (1.8,6.35);

\draw [thick] (0.5,7.5)    -- (6.0,7.5);
\draw [thick] (0.5,5.75)    -- (6.0,5.75);
\draw [thick] (0.5,5.75)    -- (0.5,7.5);
\draw [thick] (6.0,5.75)    -- (6.0,7.5);

\draw [line width=0.65mm][red] (0,0.5)    -- (1.35,1.75);
\draw [line width=0.65mm][red] (5.2,1.75)    -- (1.35,1.75);
\draw [line width=0.65mm][red] (5.2,1.75)    -- (9.5,5.6);
\draw [line width=0.65mm][red] (9.5,5.6)    -- (12.5,5.6);
\draw [line width=0.65mm][red] (12.5,5.6)    -- (15.2,8);
\draw [line width=0.5mm] [dashed][black] (3.2,0)    -- (12.2,8);

\node [left of = 1, xshift = 0.2 cm]{\LARGE{$-2$}};
\node [left of = 1, yshift = 2 cm, xshift = 0.2 cm] {\LARGE{$-1$}};
\node [left of = 1, yshift = 4 cm, xshift = 0.4 cm] {\LARGE{$0$}};
\node [left of = 1, yshift = 6 cm, xshift = 0.4 cm] {\LARGE{$1$}};
\node [left of = 1, yshift = 8 cm, xshift = 0.4 cm] {\LARGE{$2$}};

\node [below of = 1, xshift = 0 cm, yshift = 0.25cm]  {\LARGE $0.03$};
\node [below of = 1, xshift = 4 cm, yshift = 0.25cm]  {\LARGE $0.05$};
\node [below of = 1, xshift = 8 cm, yshift = 0.25cm]  {\LARGE $0.07$};
\node [below of = 1, xshift = 12 cm, yshift = 0.25cm]  {\LARGE $0.09$};
\node [below of = 1, xshift = 16 cm, yshift = 0.25cm]  {\LARGE $0.11$};
\node [right of = ymax, xshift = 2.9 cm, yshift = -1cm]   {\LARGE {$\ln(g_1(x)/g_0(x))$}};
\node [right of = ymax, xshift = 2.9 cm, yshift = -1.75cm]   {\LARGE {$\ln(p_1(x)/p_0(x))$}};
\node [below of = 1, xshift = 8 cm, yshift = -0.65cm]   {\LARGE $x$: pixel intensity};

\end{tikzpicture}}
		\caption{Likelihood ratio}
		\label{fig_llr_band}
	\end{subfigure}
	\caption{Density band model in \eqref{bounds1} with corresponding LFDs and their likelihood ratio in logarithmic scale.}  
	\label{fig_robust_density1}
\end{figure}Fig.~\ref{fig_band} also depicts the LFDs for this model, $g_0$ and $g_1$. The latter were calculated using the iterative algorithms detailed in Section~\ref{sec:problem_formulation}, using the nominal densities $p_0$ and $p_1$ as starting points. The tolerance was set to $\delta = 0.001$. By inspection, each LFD coincides with its upper or lower bound on certain intervals. Moreover, according to \eqref{eq:lrt1}, it holds that
\begin{equation}
  \frac{g_1(x)}{g_0(x)} \in \bigg\lbrace \, \frac{p_1(x)}{ p_0(x)},\, \frac{8}{25} \frac{p_1(x)}{p_0(x)},\,  \frac{25}{8} \frac{p_1(x)}{ p_0(x)},\, a_1,\, \frac{1}{a_0} \, \bigg\rbrace.
\label{eq:lrt_result1}
\end{equation}
That is, the robust likelihood ratio is a scaled or clipped version of the nominal likelihood ratio.

In Fig.~\ref{fig_llr_band}, the robust likelihood ratio is plotted in logarithmic scale. The two clipping constants are obtained as $\ln{a_1} =-1.125$ and $\ln{a_0} = - 0.8$ and are attained on the intervals $0.0367 \leq x \leq 0.0560$ and $0.0775 \leq x \leq 0.0925$, respectively. In the other ranges, the minimax test statistic is either the same as that of the parametric test (dashed line) or a scaled version. 
 
\subsection{Outlier Model}

For the second uncertainty model, we assume that the true distributions are close to the nominal ones in the sense that, under each hypothesis, a majority of the pixels is distributed according to the nominal model, whereas a fraction of $\varepsilon$ are allowed to be distributed arbitrarily. This is the classic outlier model first studied by Huber \cite{Huber1965}. For the examples in this section, we assumed a contamination ratio of \SI{40}{\percent}, i.e.,
\begin{equation}
  p_s'(x) = 0.6\, p_s(x), \quad  p_s''(x) = \infty, \quad s = 0,1.
\label{bounds2}
\end{equation}
This contamination ratio has been identified as a minimum value such that $g_0(x) \neq g_1(x)$, that is, the least favorable densities do not overlap completely. This value was also investigated for the outlier model in \cite{Pambudi2018a}. On the one hand, the assumption that only \SI{60}{\percent} of the pixels follow the nominal model is rather pessimistic. On the other hand, it is well in line with the idea of minimax robustness, i.e., preparing for the worst case. 

\begin{figure}[t]
	\centering
	\begin{subfigure}{0.9\linewidth}
		\centering
		\resizebox{0.87\linewidth}{!} {%
		\begin{tikzpicture}
		\tikzset{font={\fontsize{8pt}{11}\selectfont}}
		\begin{axis} [thin, no markers, width=9 cm, height=5.55cm, smooth, ylabel style={at={(axis description cs: 0.04,0.5)},anchor=south}, xtick={0, 0.1, 0.2, 0.3, 0.4, 0.5, 0.6, 0.7, 0.8, 0.9, 1.0},  xlabel={$x$: pixel intensity}, legend cell align={right}, grid=major, legend style={at={(0.97,0.95)},anchor=north east}, ytick={0, 2, 4, 6, 8, 10},legend entries={$g_0(x)$, $g_1(x)$}] 
		\addplot[color=brown] [thick] table [x=a, y=b, col sep=space]  {data2.csv};
		\addplot[color=blue]  [thick] table [x=a, y=c, col sep=space]  {data2.csv};
		\end{axis}
		\end{tikzpicture}
		}
		\caption{LFDs of the outlier model}
		\label{fig_lfd_clip}
	\end{subfigure}
	\begin{subfigure}{0.9\linewidth}
		\centering
		\resizebox{0.893\linewidth}{!} {\begin{tikzpicture}
\coordinate (ymax) at (0,8);
\coordinate (xmax) at (16, 0);
\coordinate (1) at (0,0);

\draw [thick] (0,0)  -- (ymax);
\draw [thick] (0,0)  -- (xmax);
\draw [thick] (16,8) -- (xmax);
\draw [thick] (16,8) -- (ymax);

\draw [thick] [lightgray](2,0) -- (2,5.75);
\draw [thick] [lightgray](2,7.5) -- (2,8);
\draw [thick] [lightgray](4,0) -- (4,5.75);
\draw [thick] [lightgray](4,7.5) -- (4,8);
\draw [thick] [lightgray](6,0) -- (6,8);
\draw [thick] [lightgray](8,0) -- (8,8);
\draw [thick] [lightgray](10,0) -- (10,8);
\draw [thick] [lightgray](12,0) -- (12,8);
\draw [thick] [lightgray](14,0) -- (14,8);

\draw [thick] [lightgray](0,2) -- (16,2);
\draw [thick] [lightgray](0,4) -- (16,4);
\draw [thick] [lightgray](5.2,6) -- (16,6);
\draw [thick] [lightgray](0,6) -- (0.5,6);

\draw [line width=0.5mm][red] (0.8,7.1)    -- (1.8,7.1);
\draw [line width=0.5mm][dashed][black]  (0.8,6.35) -- (1.8,6.35);

\draw [thick] (0.5,7.5)    -- (6.0,7.5);
\draw [thick] (0.5,5.75)    -- (6.0,5.75);
\draw [thick] (0.5,5.75)    -- (0.5,7.5);
\draw [thick] (6.0,5.75)    -- (6.0,7.5);

\draw [line width=0.65mm][red] (0,1.95)    -- (6,1.95);
\draw [line width=0.65mm][red] (6,1.95)    -- (10,6.41);
\draw [line width=0.65mm][red] (10,6.41)    -- (16,6.41);
\draw [line width=0.5mm] [dashed][black] (4.2,0)    -- (11.4,8);

\node [left of = 1, xshift = 0.2 cm]{\LARGE{$-0.4$}};
\node [left of = 1, yshift = 2 cm, xshift = 0.2 cm] {\LARGE{$-0.2$}};
\node [left of = 1, yshift = 4 cm, xshift = 0.4 cm] {\LARGE{$0$}};
\node [left of = 1, yshift = 6 cm, xshift = 0.4 cm] {\LARGE{$0.2$}};
\node [left of = 1, yshift = 8 cm, xshift = 0.4 cm] {\LARGE{$0.4$}};

\node [below of = 1, xshift = 2 cm, yshift = 0.25cm]  {\LARGE $0.06$};
\node [below of = 1, xshift = 6 cm, yshift = 0.25cm]  {\LARGE $0.065$};
\node [below of = 1, xshift = 10 cm, yshift = 0.25cm]  {\LARGE $0.07$};
\node [below of = 1, xshift = 14 cm, yshift = 0.25cm]  {\LARGE $0.075$};
\node [right of = ymax, xshift = 2.9 cm, yshift = -1cm]   {\LARGE {$\ln(g_1(x)/g_0(x))$}};
\node [right of = ymax, xshift = 2.9 cm, yshift = -1.75cm]   {\LARGE {$\ln(p_1(x)/p_0(x))$}};
\node [below of = 1, xshift = 8 cm, yshift = -0.65cm]   {\LARGE $x$: pixel intensity};

\end{tikzpicture}}
		\caption{Likelihood ratio}
		\label{fig_llr_clip}
	\end{subfigure}
	\caption{LFDs of the outlier model in \eqref{bounds2} and their likelihood ratio in logarithmic scale.} 
	\label{fig_robust_density2}
\end{figure}
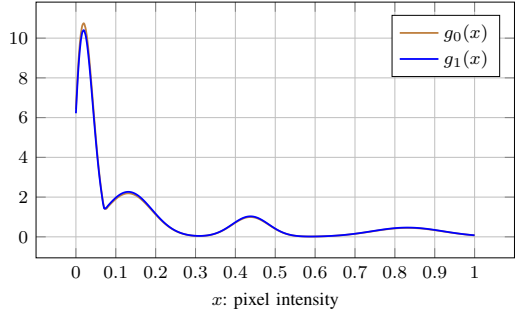
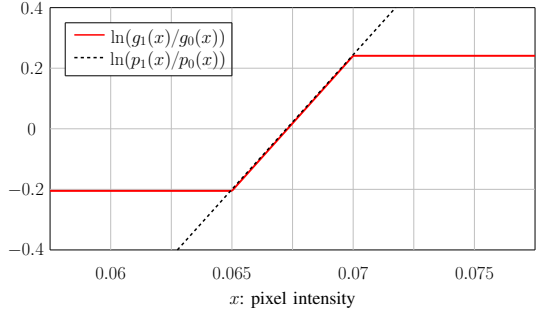

Fig.~\ref{fig_lfd_clip} shows the LFDs for the uncertainty model in \eqref{bounds1}. Note that they are difficult to distinguish by inspection since they are most similar by construction. As stated in \eqref{eq:lrt4} and shown in Fig.~\ref{fig_llr_clip}, the corresponding robust likelihood ratio in logarithmic scale satisfies
\begin{equation}
   \ln {\frac{g_1(x)}{g_0(x)}} \in \bigg\lbrace \, \ln {\frac{p_1(x)}{ p_0(x)}},\, -0.205,\, 0.241\, \bigg\rbrace. 
  \label{eq:lrt4_result}
\end{equation}
That is, it coincides with the nominal likelihood ratio on the interval $0.065 \leq x \leq 0.07$ and is clipped at either $-0.205$ or $0.241$ everywhere else. This clipping ensures that the influence of $x$ is limited and trusted only up to a certain level and illustrates a key feature of robust detectors, namely that a few outlying observations should not be able to outweigh the majority of the data. 

\section{Simulation Results}
\label{sec:simulation_results}

Multiple tomographic images obtained from different viewpoints, as described in Section II, are used for the detection. More precisely, first, the LRT is performed on each image, then the resulting binary masks are multiplied pixel-by-pixel. That is, the final target regions are given by the intersections of the target regions of the individual images. This fusion rule is also known as \emph{hard fusion} in the literature. For both the nominal and the robust test the likelihood ratio threshold $\gamma$ in \eqref{eq_lrt} is obtained by solving \eqref{eq_fa} for the estimated nominal and least favorable distributions, respectively.

\begin{figure}[t]
	\centering
	\begin{subfigure}[t]{0.875\linewidth}
		\begin{minipage}{1\linewidth}
			\resizebox{1\linewidth}{!} {\begin{tikzpicture}
\begin{overpic}[scale=0.509]{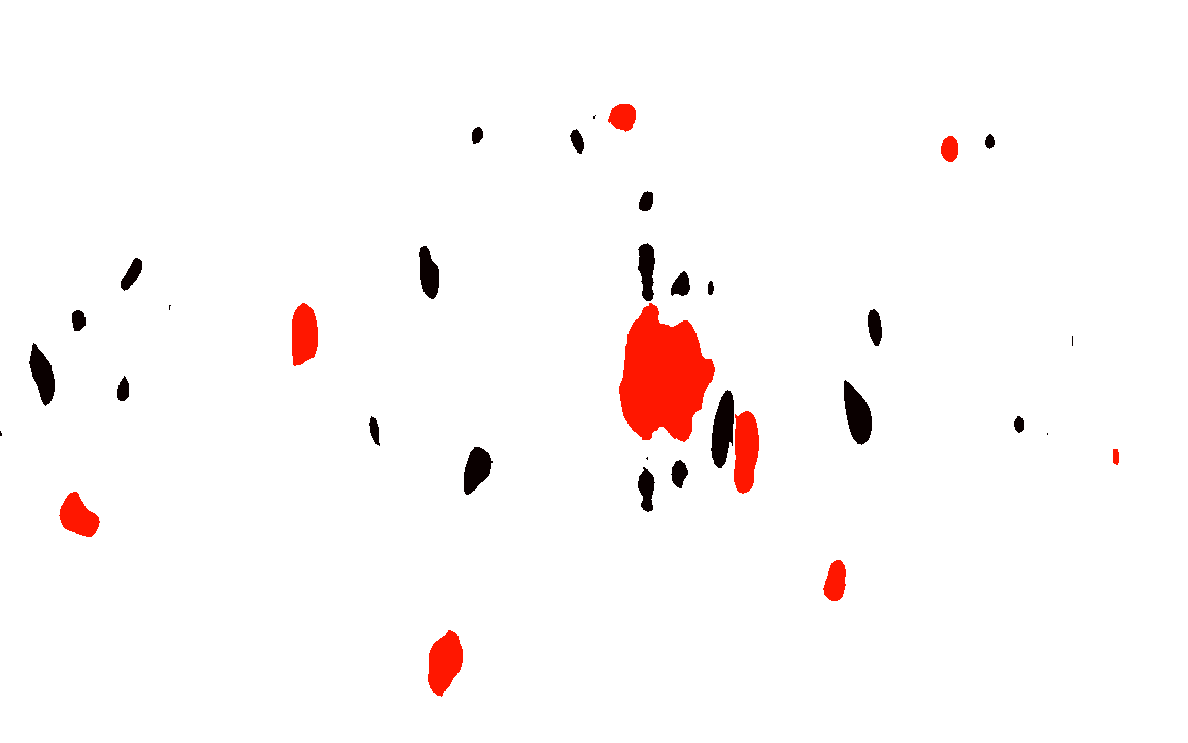}
\coordinate (ymax) at (0,10);
\coordinate (xmax) at (16, 0);
\coordinate (1) at (0,0);

\draw [thick] (0,0) -- (ymax);
\draw [thick] (0,0) -- (xmax);
\draw [thick] (16,10) -- (xmax);
\draw [thick] (16,10) -- (ymax);

\node [left of = 1, xshift = 0.3 cm]{\LARGE $-5$};
\node [left of = 1, yshift = 2 cm, xshift = 0.3 cm] {\LARGE $-3$};
\node [left of = 1, yshift = 4 cm, xshift = 0.3 cm] {\LARGE $-1$};
\node [left of = 1, yshift = 6 cm, xshift = 0.5 cm] {\LARGE $1$};
\node [left of = 1, yshift = 8 cm, xshift = 0.5 cm] {\LARGE $3$};
\node [left of = 1, yshift = 10 cm, xshift = 0.5 cm] {\LARGE $5$};
\node [left of = 1, yshift = 5 cm, xshift = -0.4 cm] {\rotatebox{90} {\LARGE y[m]}};

\draw ($(1) + (-0.2,0)$) -- ++ (0.2,0);
\draw ($(1) + (-0.2,2)$) -- ++ (0.2,0);
\draw ($(1) + (-0.2,4)$) -- ++ (0.2,0);
\draw ($(1) + (-0.2,6)$) -- ++ (0.2,0);
\draw ($(1) + (-0.2,8)$) -- ++ (0.2,0);
\draw ($(1) + (-0.2,10)$) -- ++ (0.2,0);

\node [below of = 1, xshift = 0 cm, yshift = 0.25cm]  {\LARGE$0$};
\node [below of = 1, xshift = 2 cm, yshift = 0.25cm]  {\LARGE$2$};
\node [below of = 1, xshift = 4 cm, yshift = 0.25cm]  {\LARGE$4$};
\node [below of = 1, xshift = 6 cm, yshift = 0.25cm]  {\LARGE$6$};
\node [below of = 1, xshift = 8 cm, yshift = 0.25cm]  {\LARGE$8$};
\node [below of = 1, xshift = 10 cm, yshift = 0.25cm]  {\LARGE$10$};
\node [below of = 1, xshift = 12 cm, yshift = 0.25cm]  {\LARGE$12$};
\node [below of = 1, xshift = 14 cm, yshift = 0.25cm]  {\LARGE$14$};
\node [below of = 1, xshift = 16 cm, yshift = 0.25cm]  {\LARGE$16$};
\node [below of = 1, xshift = 8 cm, yshift = -0.7cm]   {\LARGE x[m]};

\draw ($(1) + (0,-0.2)$) -- ++ (0,0.2);
\draw ($(1) + (2,-0.2)$) -- ++ (0,0.2);
\draw ($(1) + (4,-0.2)$) -- ++ (0,0.2);
\draw ($(1) + (6,-0.2)$) -- ++ (0,0.2);
\draw ($(1) + (8,-0.2)$) -- ++ (0,0.2);
\draw ($(1) + (10,-0.2)$) -- ++ (0,0.2);
\draw ($(1) + (12,-0.2)$) -- ++ (0,0.2);
\draw ($(1) + (14,-0.2)$) -- ++ (0,0.2);
\draw ($(1) + (16,-0.2)$) -- ++ (0,0.2);

\node at (1, 3) {\LARGE\color{blue}{$\times^1$}};
\node at (4, 5.5) {\LARGE\color{blue}{$\times^2$}};
\node at (6, 1) {\LARGE\color{blue}{$\times^3$}};
\node at (8.5, 8.2) {\LARGE\color{blue}{$\times^4$}};
\node at (9, 5) {\LARGE\color{blue}{$\times^5$}};
\node at (10.2, 4) {\LARGE\color{blue}{$\times^6$}};
\node at (11.5, 2) {\LARGE\color{blue}{$\times^7$}};
\node at (13, 8.2) {\LARGE\color{blue}{$\times^8$}};
\node at (15, 4) {\LARGE\color{blue}{$\times^9$}};

\end{overpic}


\end{tikzpicture}}
			\caption{Parametric LRT }
			\label{fig:bim_nom}
		\end{minipage}
	\end{subfigure}
	\begin{subfigure}[t]{0.875\linewidth}
		\begin{minipage}{1\linewidth}
			\resizebox{1\linewidth}{!} {\begin{tikzpicture}
\begin{overpic}[scale=0.4985]{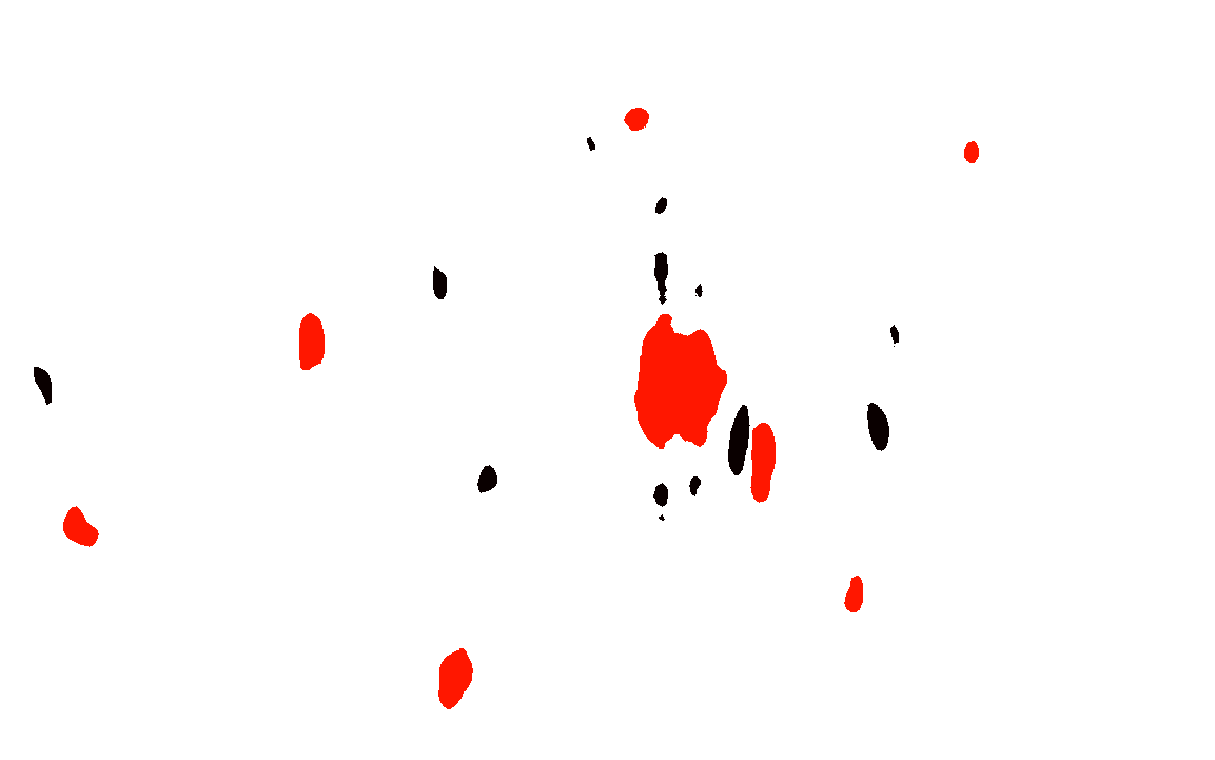}
\coordinate (ymax) at (0,10);
\coordinate (xmax) at (16, 0);
\coordinate (1) at (0,0);

\draw [thick] (0,0) -- (ymax);
\draw [thick] (0,0) -- (xmax);
\draw [thick] (16,10) -- (xmax);
\draw [thick] (16,10) -- (ymax);

\node [left of = 1, xshift = 0.3 cm]{\LARGE$-5$};
\node [left of = 1, yshift = 2 cm, xshift = 0.3 cm] {\LARGE $-3$};
\node [left of = 1, yshift = 4 cm, xshift = 0.3 cm] {\LARGE $-1$};
\node [left of = 1, yshift = 6 cm, xshift = 0.5 cm] {\LARGE $1$};
\node [left of = 1, yshift = 8 cm, xshift = 0.5 cm] {\LARGE $3$};
\node [left of = 1, yshift = 10 cm, xshift = 0.5 cm] {\LARGE $5$};
\node [left of = 1, yshift = 5 cm, xshift = -0.4 cm] {\rotatebox{90} {\LARGE y[m]}};

\draw ($(1) + (-0.2,0)$) -- ++ (0.2,0);
\draw ($(1) + (-0.2,2)$) -- ++ (0.2,0);
\draw ($(1) + (-0.2,4)$) -- ++ (0.2,0);
\draw ($(1) + (-0.2,6)$) -- ++ (0.2,0);
\draw ($(1) + (-0.2,8)$) -- ++ (0.2,0);
\draw ($(1) + (-0.2,10)$) -- ++ (0.2,0);

\node [below of = 1, xshift = 0 cm, yshift = 0.25cm]  {\LARGE$0$};
\node [below of = 1, xshift = 2 cm, yshift = 0.25cm]  {\LARGE$2$};
\node [below of = 1, xshift = 4 cm, yshift = 0.25cm]  {\LARGE$4$};
\node [below of = 1, xshift = 6 cm, yshift = 0.25cm]  {\LARGE$6$};
\node [below of = 1, xshift = 8 cm, yshift = 0.25cm]  {\LARGE$8$};
\node [below of = 1, xshift = 10 cm, yshift = 0.25cm]  {\LARGE$10$};
\node [below of = 1, xshift = 12 cm, yshift = 0.25cm]  {\LARGE$12$};
\node [below of = 1, xshift = 14 cm, yshift = 0.25cm]  {\LARGE$14$};
\node [below of = 1, xshift = 16 cm, yshift = 0.25cm]  {\LARGE$16$};
\node [below of = 1, xshift = 8 cm, yshift = -0.4cm]   {\LARGE x[m]};

\draw ($(1) + (0,-0.2)$) -- ++ (0,0.2);
\draw ($(1) + (2,-0.2)$) -- ++ (0,0.2);
\draw ($(1) + (4,-0.2)$) -- ++ (0,0.2);
\draw ($(1) + (6,-0.2)$) -- ++ (0,0.2);
\draw ($(1) + (8,-0.2)$) -- ++ (0,0.2);
\draw ($(1) + (10,-0.2)$) -- ++ (0,0.2);
\draw ($(1) + (12,-0.2)$) -- ++ (0,0.2);
\draw ($(1) + (14,-0.2)$) -- ++ (0,0.2);
\draw ($(1) + (16,-0.2)$) -- ++ (0,0.2);

\node at (1, 3) {\LARGE\color{blue}{$\times^1$}};
\node at (4, 5.5) {\LARGE\color{blue}{$\times^2$}};
\node at (6, 1) {\LARGE\color{blue}{$\times^3$}};
\node at (8.5, 8.2) {\LARGE\color{blue}{$\times^4$}};
\node at (9, 5) {\LARGE\color{blue}{$\times^5$}};
\node at (10.2, 4) {\LARGE\color{blue}{$\times^6$}};
\node at (11.5, 2) {\LARGE\color{blue}{$\times^7$}};
\node at (13, 8.2) {\LARGE\color{blue}{$\times^8$}};
\node at (15, 4) {\LARGE\color{blue}{$\times^9$}};

\end{overpic}


\end{tikzpicture}}
			\caption{Robust LRT using density band model}
			\label{fig:bim_rob1}
		\end{minipage}
	\end{subfigure}
	\begin{subfigure}[t]{0.875\linewidth}
		\begin{minipage}{1\linewidth}
			\resizebox{1\linewidth}{!} {\begin{tikzpicture}
\begin{overpic}[scale=0.509]{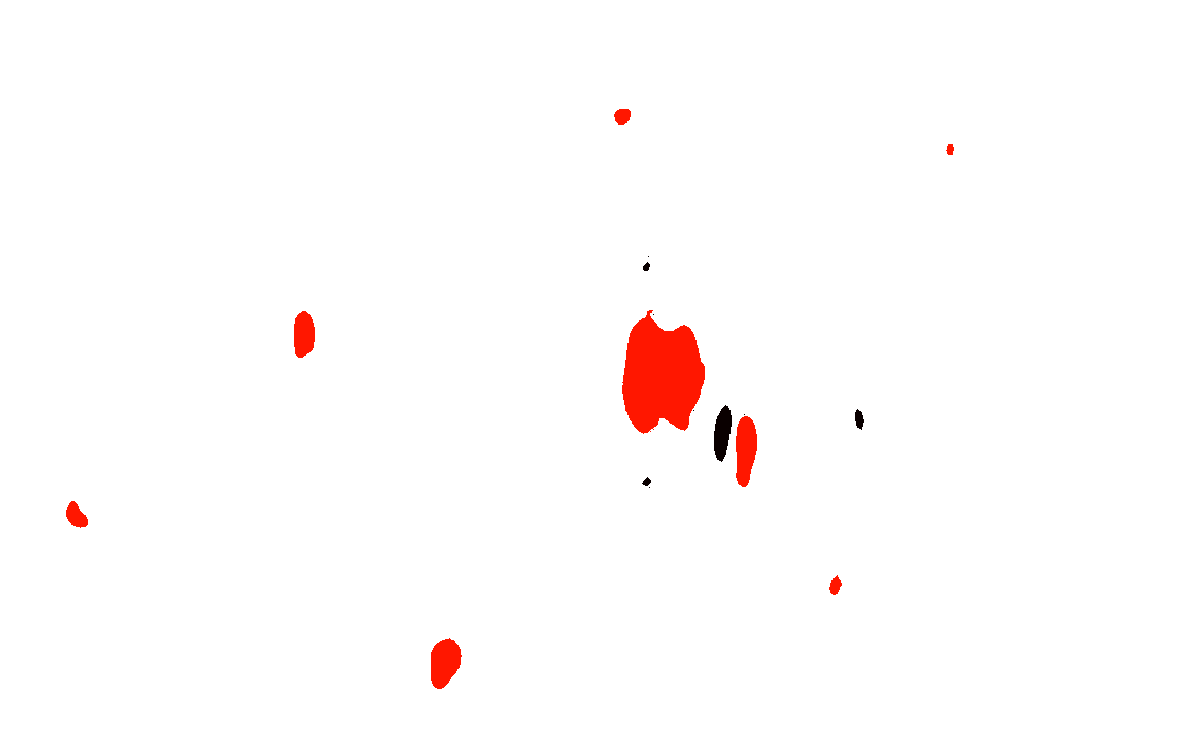}
\coordinate (ymax) at (0,10);
\coordinate (xmax) at (16, 0);
\coordinate (1) at (0,0);

\draw [thick] (0,0) -- (ymax);
\draw [thick] (0,0) -- (xmax);
\draw [thick] (16,10) -- (xmax);
\draw [thick] (16,10) -- (ymax);

\node [left of = 1, xshift = 0.3 cm]{\LARGE $-5$};
\node [left of = 1, yshift = 2 cm, xshift = 0.3 cm] {\LARGE $-3$};
\node [left of = 1, yshift = 4 cm, xshift = 0.3 cm] {\LARGE $-1$};
\node [left of = 1, yshift = 6 cm, xshift = 0.5 cm] {\LARGE $1$};
\node [left of = 1, yshift = 8 cm, xshift = 0.5 cm] {\LARGE $3$};
\node [left of = 1, yshift = 10 cm, xshift = 0.5 cm] {\LARGE $5$};
\node [left of = 1, yshift = 5 cm, xshift = -0.4cm] {\rotatebox{90} {\LARGE y[m]}};

\draw ($(1) + (-0.2,0)$) -- ++ (0.2,0);
\draw ($(1) + (-0.2,2)$) -- ++ (0.2,0);
\draw ($(1) + (-0.2,4)$) -- ++ (0.2,0);
\draw ($(1) + (-0.2,6)$) -- ++ (0.2,0);
\draw ($(1) + (-0.2,8)$) -- ++ (0.2,0);
\draw ($(1) + (-0.2,10)$) -- ++ (0.2,0);

\node [below of = 1, xshift = 0 cm, yshift = 0.25cm]  {\LARGE$0$};
\node [below of = 1, xshift = 2 cm, yshift = 0.25cm]  {\LARGE$2$};
\node [below of = 1, xshift = 4 cm, yshift = 0.25cm]  {\LARGE$4$};
\node [below of = 1, xshift = 6 cm, yshift = 0.25cm]  {\LARGE$6$};
\node [below of = 1, xshift = 8 cm, yshift = 0.25cm]  {\LARGE$8$};
\node [below of = 1, xshift = 10 cm, yshift = 0.25cm]  {\LARGE$10$};
\node [below of = 1, xshift = 12 cm, yshift = 0.25cm]  {\LARGE$12$};
\node [below of = 1, xshift = 14 cm, yshift = 0.25cm]  {\LARGE$14$};
\node [below of = 1, xshift = 16 cm, yshift = 0.25cm]  {\LARGE$16$};
\node [below of = 1, xshift = 8 cm, yshift = -0.7cm]   {\LARGE x[m]};

\draw ($(1) + (0,-0.2)$) -- ++ (0,0.2);
\draw ($(1) + (2,-0.2)$) -- ++ (0,0.2);
\draw ($(1) + (4,-0.2)$) -- ++ (0,0.2);
\draw ($(1) + (6,-0.2)$) -- ++ (0,0.2);
\draw ($(1) + (8,-0.2)$) -- ++ (0,0.2);
\draw ($(1) + (10,-0.2)$) -- ++ (0,0.2);
\draw ($(1) + (12,-0.2)$) -- ++ (0,0.2);
\draw ($(1) + (14,-0.2)$) -- ++ (0,0.2);
\draw ($(1) + (16,-0.2)$) -- ++ (0,0.2);

\node at (1, 3) {\LARGE\color{blue}{$\times^1$}};
\node at (4, 5.5) {\LARGE\color{blue}{$\times^2$}};
\node at (6, 1) {\LARGE\color{blue}{$\times^3$}};
\node at (8.5, 8.2) {\LARGE\color{blue}{$\times^4$}};
\node at (9, 5) {\LARGE\color{blue}{$\times^5$}};
\node at (10.2, 4) {\LARGE\color{blue}{$\times^6$}};
\node at (11.5, 2) {\LARGE\color{blue}{$\times^7$}};
\node at (13, 8.2) {\LARGE\color{blue}{$\times^8$}};
\node at (15, 4) {\LARGE\color{blue}{$\times^9$}};

\end{overpic}


\end{tikzpicture}}
			\caption{Robust LRT using outlier model}
			\label{fig:bim_rob2}
		\end{minipage}
	\end{subfigure}
	\caption{Detection results for $\alpha = 0.05$ in low surface roughness environments. Color coding: detected targets (red), false alarms (black).}
	\label{fig_bim}
\end{figure}
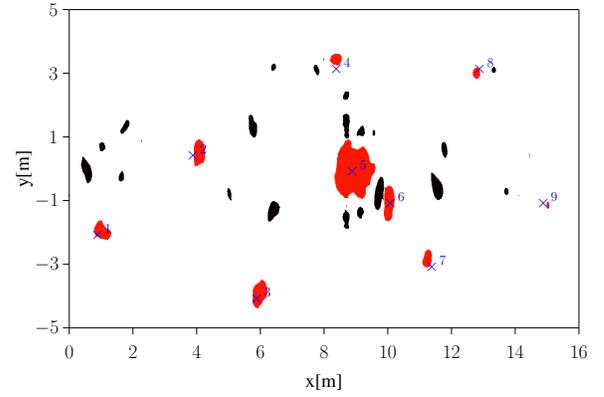
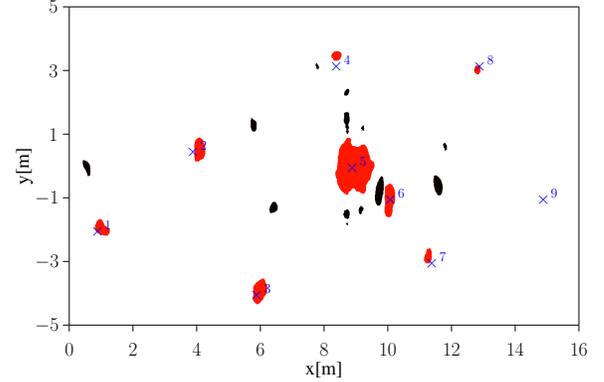
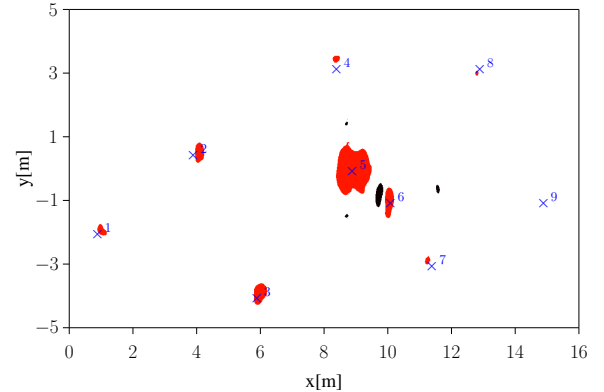

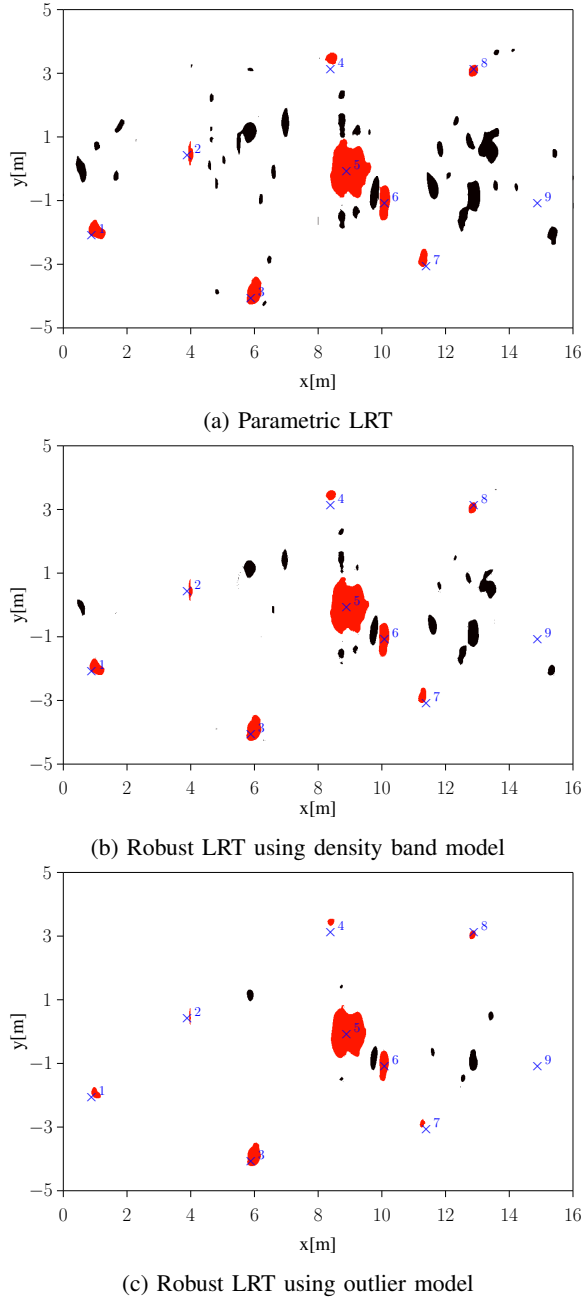
\begin{figure}[t]
	\centering
	\begin{subfigure}[t]{0.875\linewidth}
		\begin{minipage}{1\linewidth}
			\resizebox{1\linewidth}{!} {\begin{tikzpicture}
\begin{overpic}[scale=0.4985]{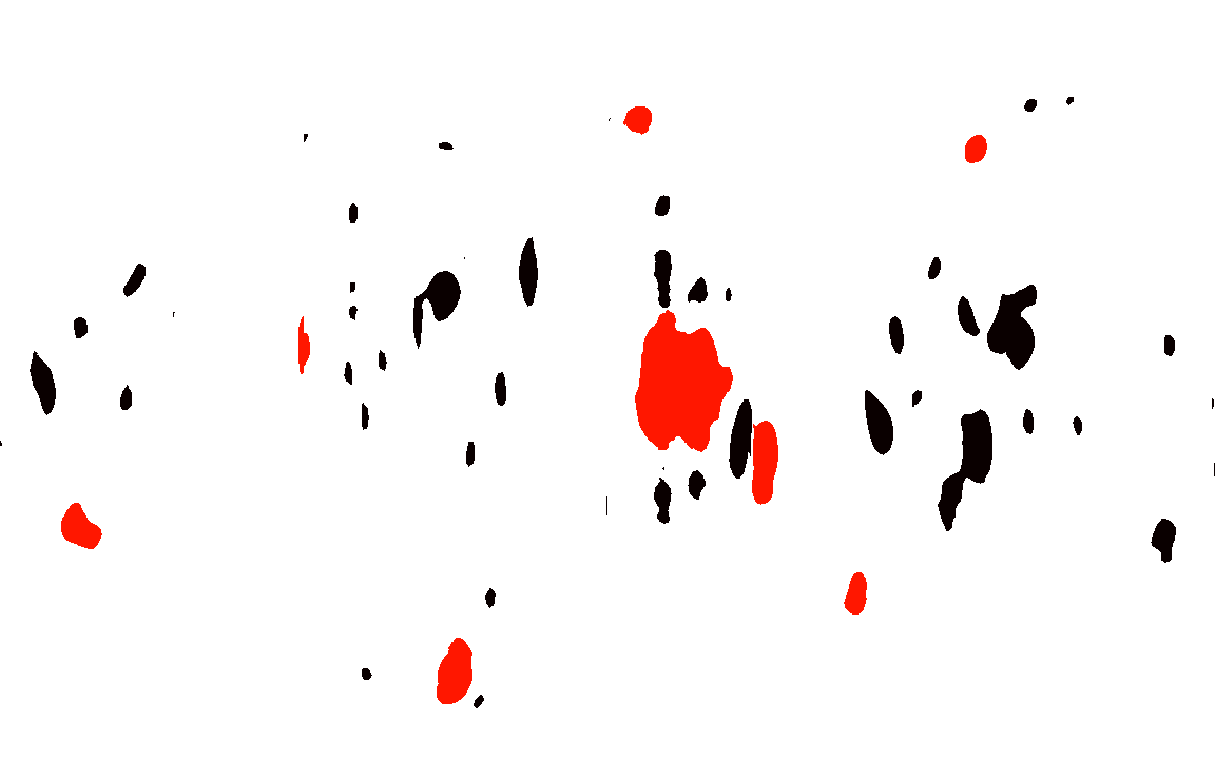}
\coordinate (ymax) at (0,10);
\coordinate (xmax) at (16, 0);
\coordinate (1) at (0,0);

\draw [thick] (0,0) -- (ymax);
\draw [thick] (0,0) -- (xmax);
\draw [thick] (16,10) -- (xmax);
\draw [thick] (16,10) -- (ymax);

\node [left of = 1, xshift = 0.3 cm]{\LARGE $-5$};
\node [left of = 1, yshift = 2 cm, xshift = 0.3 cm] {\LARGE $-3$};
\node [left of = 1, yshift = 4 cm, xshift = 0.3 cm] {\LARGE $-1$};
\node [left of = 1, yshift = 6 cm, xshift = 0.5 cm] {\LARGE $1$};
\node [left of = 1, yshift = 8 cm, xshift = 0.5 cm] {\LARGE $3$};
\node [left of = 1, yshift = 10 cm, xshift = 0.5 cm] {\LARGE $5$};
\node [left of = 1, yshift = 5 cm, xshift = -0.4 cm] {\rotatebox{90} {\LARGE y[m]}};

\draw ($(1) + (-0.2,0)$) -- ++ (0.2,0);
\draw ($(1) + (-0.2,2)$) -- ++ (0.2,0);
\draw ($(1) + (-0.2,4)$) -- ++ (0.2,0);
\draw ($(1) + (-0.2,6)$) -- ++ (0.2,0);
\draw ($(1) + (-0.2,8)$) -- ++ (0.2,0);
\draw ($(1) + (-0.2,10)$) -- ++ (0.2,0);

\node [below of = 1, xshift = 0 cm, yshift = 0.25cm]  {\LARGE$0$};
\node [below of = 1, xshift = 2 cm, yshift = 0.25cm]  {\LARGE$2$};
\node [below of = 1, xshift = 4 cm, yshift = 0.25cm]  {\LARGE$4$};
\node [below of = 1, xshift = 6 cm, yshift = 0.25cm]  {\LARGE$6$};
\node [below of = 1, xshift = 8 cm, yshift = 0.25cm]  {\LARGE$8$};
\node [below of = 1, xshift = 10 cm, yshift = 0.25cm]  {\LARGE$10$};
\node [below of = 1, xshift = 12 cm, yshift = 0.25cm]  {\LARGE$12$};
\node [below of = 1, xshift = 14 cm, yshift = 0.25cm]  {\LARGE$14$};
\node [below of = 1, xshift = 16 cm, yshift = 0.25cm]  {\LARGE$16$};
\node [below of = 1, xshift = 8 cm, yshift = -0.7cm]   {\LARGE x[m]};

\draw ($(1) + (0,-0.2)$) -- ++ (0,0.2);
\draw ($(1) + (2,-0.2)$) -- ++ (0,0.2);
\draw ($(1) + (4,-0.2)$) -- ++ (0,0.2);
\draw ($(1) + (6,-0.2)$) -- ++ (0,0.2);
\draw ($(1) + (8,-0.2)$) -- ++ (0,0.2);
\draw ($(1) + (10,-0.2)$) -- ++ (0,0.2);
\draw ($(1) + (12,-0.2)$) -- ++ (0,0.2);
\draw ($(1) + (14,-0.2)$) -- ++ (0,0.2);
\draw ($(1) + (16,-0.2)$) -- ++ (0,0.2);

\node at (1, 3) {\LARGE\color{blue}{$\times^1$}};
\node at (4, 5.5) {\LARGE\color{blue}{$\times^2$}};
\node at (6, 1) {\LARGE\color{blue}{$\times^3$}};
\node at (8.5, 8.2) {\LARGE\color{blue}{$\times^4$}};
\node at (9, 5) {\LARGE\color{blue}{$\times^5$}};
\node at (10.2, 4) {\LARGE\color{blue}{$\times^6$}};
\node at (11.5, 2) {\LARGE\color{blue}{$\times^7$}};
\node at (13, 8.2) {\LARGE\color{blue}{$\times^8$}};
\node at (15, 4) {\LARGE\color{blue}{$\times^9$}};

\end{overpic}


\end{tikzpicture}}
			\caption{Parametric LRT}
			\label{fig_nom}
		\end{minipage}
	\end{subfigure}
	\begin{subfigure}[t]{0.875\linewidth}
		\begin{minipage}{1\linewidth}
			\resizebox{1\linewidth}{!} {\begin{tikzpicture}
\begin{overpic}[scale=0.509]{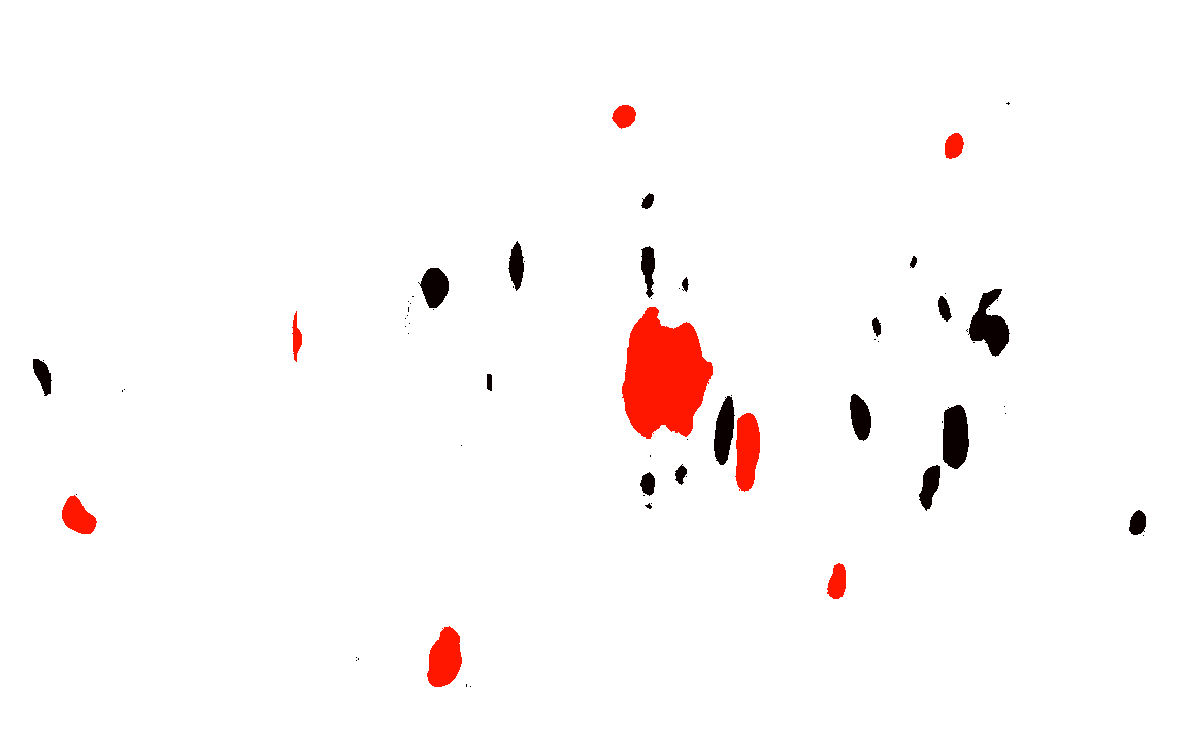}
\coordinate (ymax) at (0,10);
\coordinate (xmax) at (16, 0);
\coordinate (1) at (0,0);

\draw [thick] (0,0) -- (ymax);
\draw [thick] (0,0) -- (xmax);
\draw [thick] (16,10) -- (xmax);
\draw [thick] (16,10) -- (ymax);

\node [left of = 1, xshift = 0.3 cm]{\LARGE $-5$};
\node [left of = 1, yshift = 2 cm, xshift = 0.3 cm] {\LARGE $-3$};
\node [left of = 1, yshift = 4 cm, xshift = 0.3 cm] {\LARGE $-1$};
\node [left of = 1, yshift = 6 cm, xshift = 0.5 cm] {\LARGE $1$};
\node [left of = 1, yshift = 8 cm, xshift = 0.5 cm] {\LARGE $3$};
\node [left of = 1, yshift = 10 cm, xshift = 0.5 cm] {\LARGE $5$};
\node [left of = 1, yshift = 5 cm, xshift = -0.4 cm] {\rotatebox{90} {\LARGE y[m]}};

\draw ($(1) + (-0.2,0)$) -- ++ (0.2,0);
\draw ($(1) + (-0.2,2)$) -- ++ (0.2,0);
\draw ($(1) + (-0.2,4)$) -- ++ (0.2,0);
\draw ($(1) + (-0.2,6)$) -- ++ (0.2,0);
\draw ($(1) + (-0.2,8)$) -- ++ (0.2,0);
\draw ($(1) + (-0.2,10)$) -- ++ (0.2,0);

\node [below of = 1, xshift = 0 cm, yshift = 0.25cm]  {\LARGE$0$};
\node [below of = 1, xshift = 2 cm, yshift = 0.25cm]  {\LARGE$2$};
\node [below of = 1, xshift = 4 cm, yshift = 0.25cm]  {\LARGE$4$};
\node [below of = 1, xshift = 6 cm, yshift = 0.25cm]  {\LARGE$6$};
\node [below of = 1, xshift = 8 cm, yshift = 0.25cm]  {\LARGE$8$};
\node [below of = 1, xshift = 10 cm, yshift = 0.25cm]  {\LARGE$10$};
\node [below of = 1, xshift = 12 cm, yshift = 0.25cm]  {\LARGE$12$};
\node [below of = 1, xshift = 14 cm, yshift = 0.25cm]  {\LARGE$14$};
\node [below of = 1, xshift = 16 cm, yshift = 0.25cm]  {\LARGE$16$};
\node [below of = 1, xshift = 8 cm, yshift = -0.4cm]   {\LARGE x[m]};

\draw ($(1) + (0,-0.2)$) -- ++ (0,0.2);
\draw ($(1) + (2,-0.2)$) -- ++ (0,0.2);
\draw ($(1) + (4,-0.2)$) -- ++ (0,0.2);
\draw ($(1) + (6,-0.2)$) -- ++ (0,0.2);
\draw ($(1) + (8,-0.2)$) -- ++ (0,0.2);
\draw ($(1) + (10,-0.2)$) -- ++ (0,0.2);
\draw ($(1) + (12,-0.2)$) -- ++ (0,0.2);
\draw ($(1) + (14,-0.2)$) -- ++ (0,0.2);
\draw ($(1) + (16,-0.2)$) -- ++ (0,0.2);

\node at (1, 3) {\LARGE\color{blue}{$\times^1$}};
\node at (4, 5.5) {\LARGE\color{blue}{$\times^2$}};
\node at (6, 1) {\LARGE\color{blue}{$\times^3$}};
\node at (8.5, 8.2) {\LARGE\color{blue}{$\times^4$}};
\node at (9, 5) {\LARGE\color{blue}{$\times^5$}};
\node at (10.2, 4) {\LARGE\color{blue}{$\times^6$}};
\node at (11.5, 2) {\LARGE\color{blue}{$\times^7$}};
\node at (13, 8.2) {\LARGE\color{blue}{$\times^8$}};
\node at (15, 4) {\LARGE\color{blue}{$\times^9$}};

\end{overpic}


\end{tikzpicture}}
			\caption{Robust LRT using density band model}
			\label{fig_rob1}
		\end{minipage}
	\end{subfigure}
	\begin{subfigure}[t]{0.875\linewidth}
		\begin{minipage}{1\linewidth}
			\resizebox{1\linewidth}{!} {\begin{tikzpicture}
\begin{overpic}[scale=0.509]{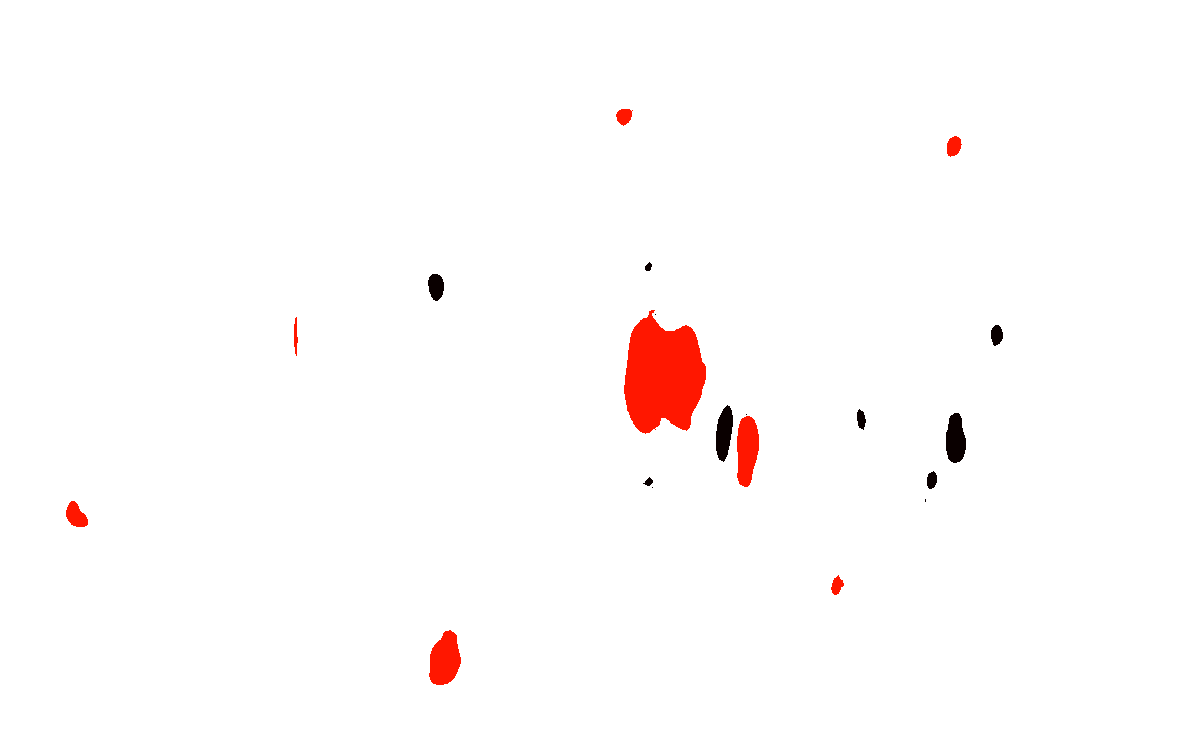}
\coordinate (ymax) at (0,10);
\coordinate (xmax) at (16, 0);
\coordinate (1) at (0,0);

\draw [thick] (0,0) -- (ymax);
\draw [thick] (0,0) -- (xmax);
\draw [thick] (16,10) -- (xmax);
\draw [thick] (16,10) -- (ymax);

\node [left of = 1, xshift = 0.3 cm]{\LARGE $-5$};
\node [left of = 1, yshift = 2 cm, xshift = 0.3 cm] {\LARGE $-3$};
\node [left of = 1, yshift = 4 cm, xshift = 0.3 cm] {\LARGE $-1$};
\node [left of = 1, yshift = 6 cm, xshift = 0.5 cm] {\LARGE $1$};
\node [left of = 1, yshift = 8 cm, xshift = 0.5 cm] {\LARGE $3$};
\node [left of = 1, yshift = 10 cm, xshift = 0.5 cm] {\LARGE $5$};
\node [left of = 1, yshift = 5 cm, xshift = -0.4cm] {\rotatebox{90} {\LARGE y[m]}};

\draw ($(1) + (-0.2,0)$) -- ++ (0.2,0);
\draw ($(1) + (-0.2,2)$) -- ++ (0.2,0);
\draw ($(1) + (-0.2,4)$) -- ++ (0.2,0);
\draw ($(1) + (-0.2,6)$) -- ++ (0.2,0);
\draw ($(1) + (-0.2,8)$) -- ++ (0.2,0);
\draw ($(1) + (-0.2,10)$) -- ++ (0.2,0);

\node [below of = 1, xshift = 0 cm, yshift = 0.25cm]  {\LARGE$0$};
\node [below of = 1, xshift = 2 cm, yshift = 0.25cm]  {\LARGE$2$};
\node [below of = 1, xshift = 4 cm, yshift = 0.25cm]  {\LARGE$4$};
\node [below of = 1, xshift = 6 cm, yshift = 0.25cm]  {\LARGE$6$};
\node [below of = 1, xshift = 8 cm, yshift = 0.25cm]  {\LARGE$8$};
\node [below of = 1, xshift = 10 cm, yshift = 0.25cm]  {\LARGE$10$};
\node [below of = 1, xshift = 12 cm, yshift = 0.25cm]  {\LARGE$12$};
\node [below of = 1, xshift = 14 cm, yshift = 0.25cm]  {\LARGE$14$};
\node [below of = 1, xshift = 16 cm, yshift = 0.25cm]  {\LARGE$16$};
\node [below of = 1, xshift = 8 cm, yshift = -0.7cm]   {\LARGE x[m]};

\draw ($(1) + (0,-0.2)$) -- ++ (0,0.2);
\draw ($(1) + (2,-0.2)$) -- ++ (0,0.2);
\draw ($(1) + (4,-0.2)$) -- ++ (0,0.2);
\draw ($(1) + (6,-0.2)$) -- ++ (0,0.2);
\draw ($(1) + (8,-0.2)$) -- ++ (0,0.2);
\draw ($(1) + (10,-0.2)$) -- ++ (0,0.2);
\draw ($(1) + (12,-0.2)$) -- ++ (0,0.2);
\draw ($(1) + (14,-0.2)$) -- ++ (0,0.2);
\draw ($(1) + (16,-0.2)$) -- ++ (0,0.2);

\node at (1, 3) {\LARGE\color{blue}{$\times^1$}};
\node at (4, 5.5) {\LARGE\color{blue}{$\times^2$}};
\node at (6, 1) {\LARGE\color{blue}{$\times^3$}};
\node at (8.5, 8.2) {\LARGE\color{blue}{$\times^4$}};
\node at (9, 5) {\LARGE\color{blue}{$\times^5$}};
\node at (10.2, 4) {\LARGE\color{blue}{$\times^6$}};
\node at (11.5, 2) {\LARGE\color{blue}{$\times^7$}};
\node at (13, 8.2) {\LARGE\color{blue}{$\times^8$}};
\node at (15, 4) {\LARGE\color{blue}{$\times^9$}};

\end{overpic}


\end{tikzpicture}}
			\caption{Robust LRT using outlier model}
			\label{fig_rob2}
		\end{minipage}
	\end{subfigure}
	\caption{Detection results for $\alpha = 0.05$ in mixed surface roughness environments. Color coding: detected targets (red), false alarms (black).}
	\label{fig_bim2}
\end{figure}

\subsection{Detection Results in Low Surface Roughness Environments}

The binary image resulting from the parametric LRT is shown in Fig.~\ref{fig:bim_nom} for a targeted false alarm rate of $\alpha = 0.05$. The blue crosses indicate the central position of the true targets. White pixels indicate a correct decision for $H_0$ (no target present), red pixels indicate a correct decision for $H_1$ (target present), and black pixels indicate an erroneous decision for $H_1$ (false alarm). As can be seen, all nine targets are successfully detected, but there are several false alarms as well. The large red area at \SI{9}{\meter} downrange is due to the strong reflection from the metallic AT landmine placed on the surface. 

Fig.~\ref{fig:bim_rob1} depicts the result of the minimax LRT based on the density band model. Here, the number of false alarms is reduced significantly at the cost of one missing target, which is a buried plastic AT landmine located at the farthest downrange. This result is to be expected: Since the robust detector uses much weaker assumptions about the clutter distribution, it is less likely to confuse strong clutter echoes for targets. On the downside, this also makes the detector less sensitive towards weak target echoes. 

It should be highlighted that the detectors whose results are shown in Fig.~\ref{fig:bim_nom} and Fig.~\ref{fig:bim_rob1} are both designed under the same false alarm constraint.\footnote{Determining whether or not the targeted false alarm rate is met is non-trivial since the binary images also depend on the region growing algorithm, whose effects are not taken into account in the design of the detector.} Moreover, the effect of using least favorable distributions is distinctly different from simply increasing the detection threshold of the parametric test. This can be seen from the fact that the target regions for all detected landmines are approximately of the same size. Increasing the threshold of the nominal test would also reduce the number of false alarms, but at the cost of significantly smaller target regions.

This effect can indeed be observed in Fig.~\ref{fig:bim_rob2}, where the detection results of the minimax detector based on the second uncertainty model, the outlier model in \eqref{bounds1}, are depicted. The outlier model allows for significantly more uncertainty than the band model so that it is even less sensitive to clutter, while still correctly detecting all targets except for the plastic landmine 9. Hence, in this examples the outlier model can be argued to yield the best results. However, one should note that under the larger uncertainty model, the correctly detected targets are starting to be affected in the sense that the corresponding target regions are noticeably smaller and some landmines are ``almost'' missed. Given the grave consequences of having missed a target, this trade-off needs to be kept in mind when designing robust tests for landmine detection.

\subsection{Detection Results in Mixed Surface Roughness Environments}

A main advantage of minimax robust detectors is that they work well under various conditions, without having to estimate the corresponding parameters. However, this also implies that robust detectors do not adapt to the observed data. Consequently, the advantages of robust detectors should be most pronounced in scenarios where adapting to the environment is difficult, i.e., the true distributions cannot reliably be estimated from the available training data. In FL-GPR, such situations occur when the statistical properties of the clutter are different for each measurement. This can happen, for example, in environments where the surface roughness of the ground changes frequently. 

In order to simulate such a scenario, the region of interest shown in Fig.~\ref{fig_measurement} is divided into four areas whose roughness alternates between the low and the high profile. Each area has a size of \SI{4}{\meter} $\times$ \SI{10}{\meter}. Areas in the downrange from \SI{4}{\meter} to \SI{8}{\meter} and from \SI{12}{\meter} to \SI{16}{\meter} admit a high roughness profile ($h_{rms} = \SI{1.6}{\centi\meter}$, $l_c = \SI{14.93}{\centi\meter}$), while the remaining areas admit the low roughness profile ($h_{rms} = \SI{0.8}{\centi\meter}$, $l_c = \SI{14.26}{\centi\meter}$) that was used to generate the training data. The tomographic image for the higher roughness profile is shown in Fig.~\ref{fig_tomb}. The detection results for this scenario are shown in Fig.~\ref{fig_bim2}. All detectors generally exhibit an increased number of errors, especially on the segments of high roughness. However, it can be seen that the robust tests are significantly less affected by the changing clutter properties. Especially the one based on the outlier uncertainty model manages to keep the number of false alarms in check, without increasing the number of missed targets.

\subsection{Detection Results in High Surface Roughness Environments}

\begin{figure}[t]
	\centering
	\begin{subfigure}[t]{0.875\linewidth}
		\begin{minipage}{1\linewidth}
			\resizebox{1\linewidth}{!} {\begin{tikzpicture}
\begin{overpic}[scale=0.4985]{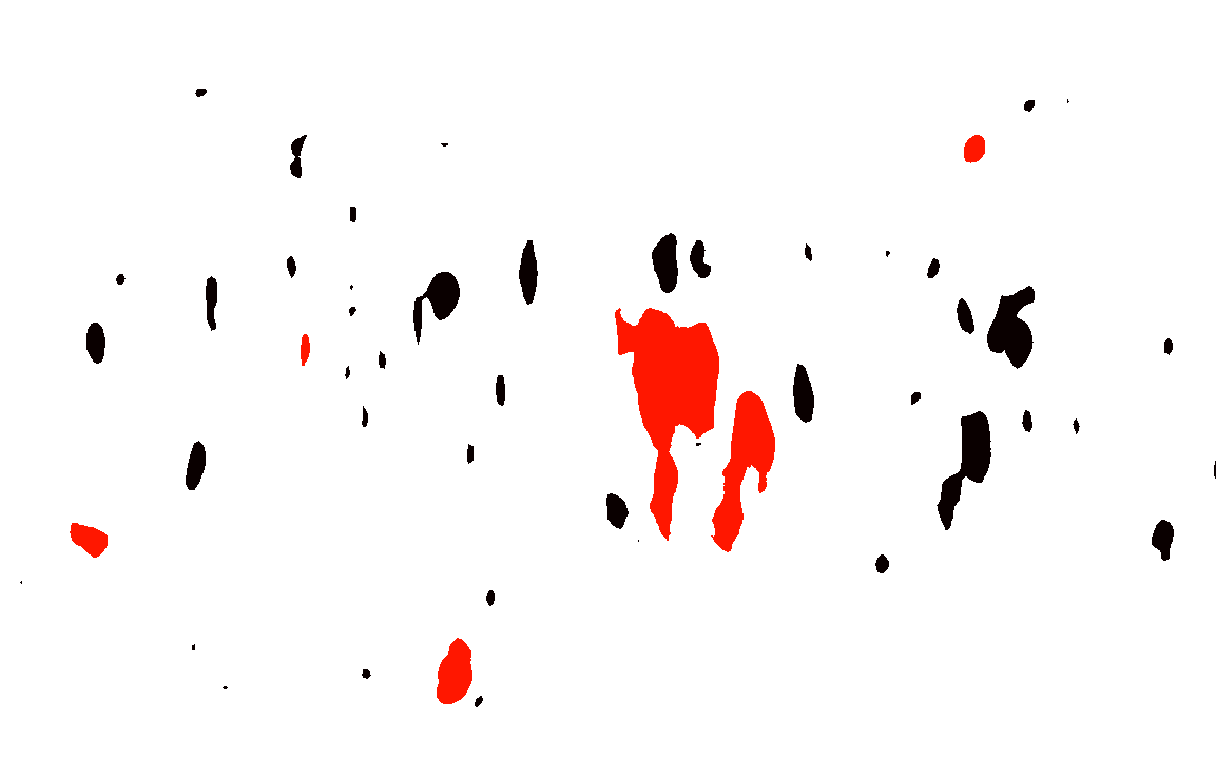}
\coordinate (ymax) at (0,10);
\coordinate (xmax) at (16, 0);
\coordinate (1) at (0,0);

\draw [thick] (0,0) -- (ymax);
\draw [thick] (0,0) -- (xmax);
\draw [thick] (16,10) -- (xmax);
\draw [thick] (16,10) -- (ymax);

\node [left of = 1, xshift = 0.3 cm]{\LARGE $-5$};
\node [left of = 1, yshift = 2 cm, xshift = 0.3 cm] {\LARGE $-3$};
\node [left of = 1, yshift = 4 cm, xshift = 0.3 cm] {\LARGE $-1$};
\node [left of = 1, yshift = 6 cm, xshift = 0.5 cm] {\LARGE $1$};
\node [left of = 1, yshift = 8 cm, xshift = 0.5 cm] {\LARGE $3$};
\node [left of = 1, yshift = 10 cm, xshift = 0.5 cm] {\LARGE $5$};
\node [left of = 1, yshift = 5 cm, xshift = -0.4 cm] {\rotatebox{90} {\LARGE y[m]}};

\draw ($(1) + (-0.2,0)$) -- ++ (0.2,0);
\draw ($(1) + (-0.2,2)$) -- ++ (0.2,0);
\draw ($(1) + (-0.2,4)$) -- ++ (0.2,0);
\draw ($(1) + (-0.2,6)$) -- ++ (0.2,0);
\draw ($(1) + (-0.2,8)$) -- ++ (0.2,0);
\draw ($(1) + (-0.2,10)$) -- ++ (0.2,0);

\node [below of = 1, xshift = 0 cm, yshift = 0.25cm]  {\LARGE$0$};
\node [below of = 1, xshift = 2 cm, yshift = 0.25cm]  {\LARGE$2$};
\node [below of = 1, xshift = 4 cm, yshift = 0.25cm]  {\LARGE$4$};
\node [below of = 1, xshift = 6 cm, yshift = 0.25cm]  {\LARGE$6$};
\node [below of = 1, xshift = 8 cm, yshift = 0.25cm]  {\LARGE$8$};
\node [below of = 1, xshift = 10 cm, yshift = 0.25cm]  {\LARGE$10$};
\node [below of = 1, xshift = 12 cm, yshift = 0.25cm]  {\LARGE$12$};
\node [below of = 1, xshift = 14 cm, yshift = 0.25cm]  {\LARGE$14$};
\node [below of = 1, xshift = 16 cm, yshift = 0.25cm]  {\LARGE$16$};
\node [below of = 1, xshift = 8 cm, yshift = -0.7cm]   {\LARGE x[m]};

\draw ($(1) + (0,-0.2)$) -- ++ (0,0.2);
\draw ($(1) + (2,-0.2)$) -- ++ (0,0.2);
\draw ($(1) + (4,-0.2)$) -- ++ (0,0.2);
\draw ($(1) + (6,-0.2)$) -- ++ (0,0.2);
\draw ($(1) + (8,-0.2)$) -- ++ (0,0.2);
\draw ($(1) + (10,-0.2)$) -- ++ (0,0.2);
\draw ($(1) + (12,-0.2)$) -- ++ (0,0.2);
\draw ($(1) + (14,-0.2)$) -- ++ (0,0.2);
\draw ($(1) + (16,-0.2)$) -- ++ (0,0.2);

\node at (1, 3) {\LARGE\color{blue}{$\times^1$}};
\node at (4, 5.5) {\LARGE\color{blue}{$\times^2$}};
\node at (6, 1) {\LARGE\color{blue}{$\times^3$}};
\node at (8.5, 8.2) {\LARGE\color{blue}{$\times^4$}};
\node at (9, 5) {\LARGE\color{blue}{$\times^5$}};
\node at (10, 4) {\LARGE\color{blue}{$\times^6$}};
\node at (11.5, 2) {\LARGE\color{blue}{$\times^7$}};
\node at (13, 8.2) {\LARGE\color{blue}{$\times^8$}};
\node at (15, 4) {\LARGE\color{blue}{$\times^9$}};

\end{overpic}


\end{tikzpicture}}
			\caption{Parametric LRT}
			\label{fig_nom_fullmiss}
		\end{minipage}
	\end{subfigure}
	\begin{subfigure}[t]{0.875\linewidth}
		\begin{minipage}{1\linewidth}
			\resizebox{1\linewidth}{!} {\begin{tikzpicture}
\begin{overpic}[scale=0.4985]{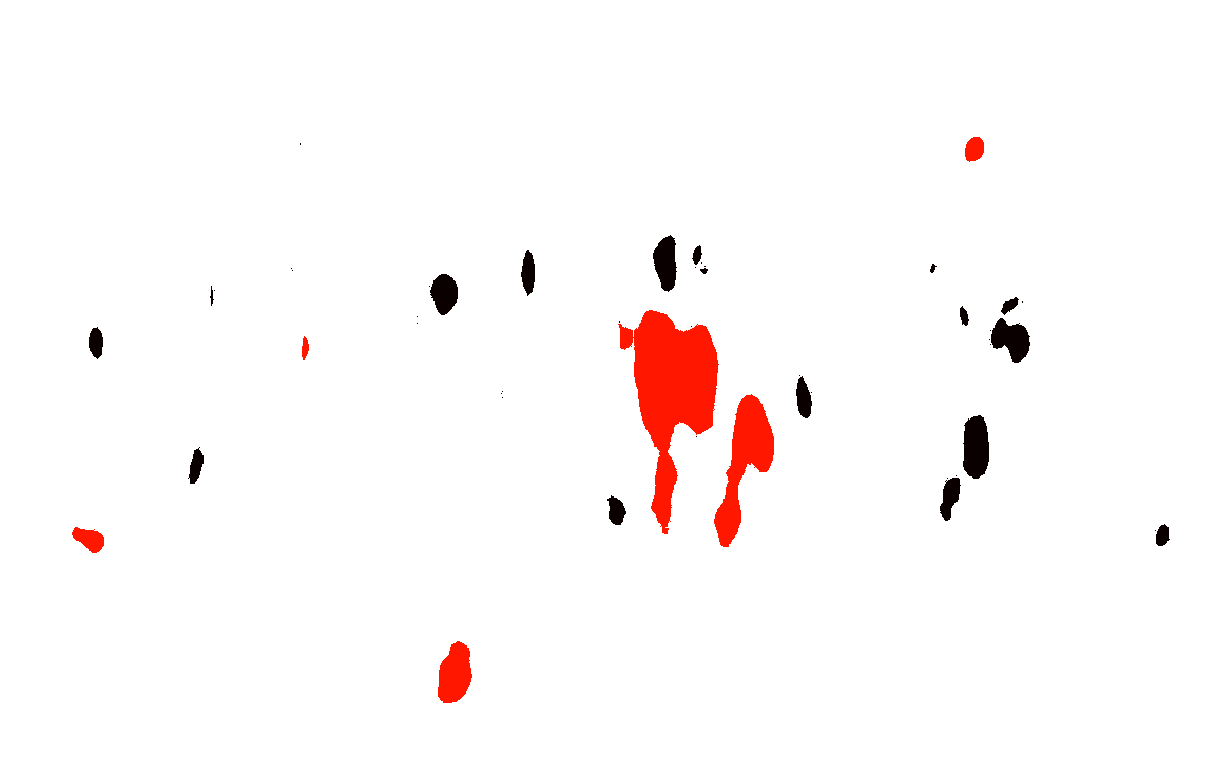}
\coordinate (ymax) at (0,10);
\coordinate (xmax) at (16, 0);
\coordinate (1) at (0,0);

\draw [thick] (0,0) -- (ymax);
\draw [thick] (0,0) -- (xmax);
\draw [thick] (16,10) -- (xmax);
\draw [thick] (16,10) -- (ymax);

\node [left of = 1, xshift = 0.3 cm]{\LARGE $-5$};
\node [left of = 1, yshift = 2 cm, xshift = 0.3 cm] {\LARGE $-3$};
\node [left of = 1, yshift = 4 cm, xshift = 0.3 cm] {\LARGE $-1$};
\node [left of = 1, yshift = 6 cm, xshift = 0.5 cm] {\LARGE $1$};
\node [left of = 1, yshift = 8 cm, xshift = 0.5 cm] {\LARGE $3$};
\node [left of = 1, yshift = 10 cm, xshift = 0.5 cm] {\LARGE $5$};
\node [left of = 1, yshift = 5 cm, xshift = -0.4 cm] {\rotatebox{90} {\LARGE y[m]}};

\draw ($(1) + (-0.2,0)$) -- ++ (0.2,0);
\draw ($(1) + (-0.2,2)$) -- ++ (0.2,0);
\draw ($(1) + (-0.2,4)$) -- ++ (0.2,0);
\draw ($(1) + (-0.2,6)$) -- ++ (0.2,0);
\draw ($(1) + (-0.2,8)$) -- ++ (0.2,0);
\draw ($(1) + (-0.2,10)$) -- ++ (0.2,0);

\node [below of = 1, xshift = 0 cm, yshift = 0.25cm]  {\LARGE$0$};
\node [below of = 1, xshift = 2 cm, yshift = 0.25cm]  {\LARGE$2$};
\node [below of = 1, xshift = 4 cm, yshift = 0.25cm]  {\LARGE$4$};
\node [below of = 1, xshift = 6 cm, yshift = 0.25cm]  {\LARGE$6$};
\node [below of = 1, xshift = 8 cm, yshift = 0.25cm]  {\LARGE$8$};
\node [below of = 1, xshift = 10 cm, yshift = 0.25cm]  {\LARGE$10$};
\node [below of = 1, xshift = 12 cm, yshift = 0.25cm]  {\LARGE$12$};
\node [below of = 1, xshift = 14 cm, yshift = 0.25cm]  {\LARGE$14$};
\node [below of = 1, xshift = 16 cm, yshift = 0.25cm]  {\LARGE$16$};
\node [below of = 1, xshift = 8 cm, yshift = -0.4cm]   {\LARGE x[m]};

\draw ($(1) + (0,-0.2)$) -- ++ (0,0.2);
\draw ($(1) + (2,-0.2)$) -- ++ (0,0.2);
\draw ($(1) + (4,-0.2)$) -- ++ (0,0.2);
\draw ($(1) + (6,-0.2)$) -- ++ (0,0.2);
\draw ($(1) + (8,-0.2)$) -- ++ (0,0.2);
\draw ($(1) + (10,-0.2)$) -- ++ (0,0.2);
\draw ($(1) + (12,-0.2)$) -- ++ (0,0.2);
\draw ($(1) + (14,-0.2)$) -- ++ (0,0.2);
\draw ($(1) + (16,-0.2)$) -- ++ (0,0.2);

\node at (1, 3) {\LARGE\color{blue}{$\times^1$}};
\node at (4, 5.5) {\LARGE\color{blue}{$\times^2$}};
\node at (6, 1) {\LARGE\color{blue}{$\times^3$}};
\node at (8.5, 8.2) {\LARGE\color{blue}{$\times^4$}};
\node at (9, 5) {\LARGE\color{blue}{$\times^5$}};
\node at (10, 4) {\LARGE\color{blue}{$\times^6$}};
\node at (11.5, 2) {\LARGE\color{blue}{$\times^7$}};
\node at (13, 8.2) {\LARGE\color{blue}{$\times^8$}};
\node at (15, 4) {\LARGE\color{blue}{$\times^9$}};

\end{overpic}


\end{tikzpicture}}
			\caption{Robust LRT using density band model}
			\label{fig_rob1_fullmiss}
		\end{minipage}
	\end{subfigure}
	\begin{subfigure}[t]{0.875\linewidth}
		\begin{minipage}{1\linewidth}
			\resizebox{1\linewidth}{!} {\begin{tikzpicture}
\begin{overpic}[scale=0.509]{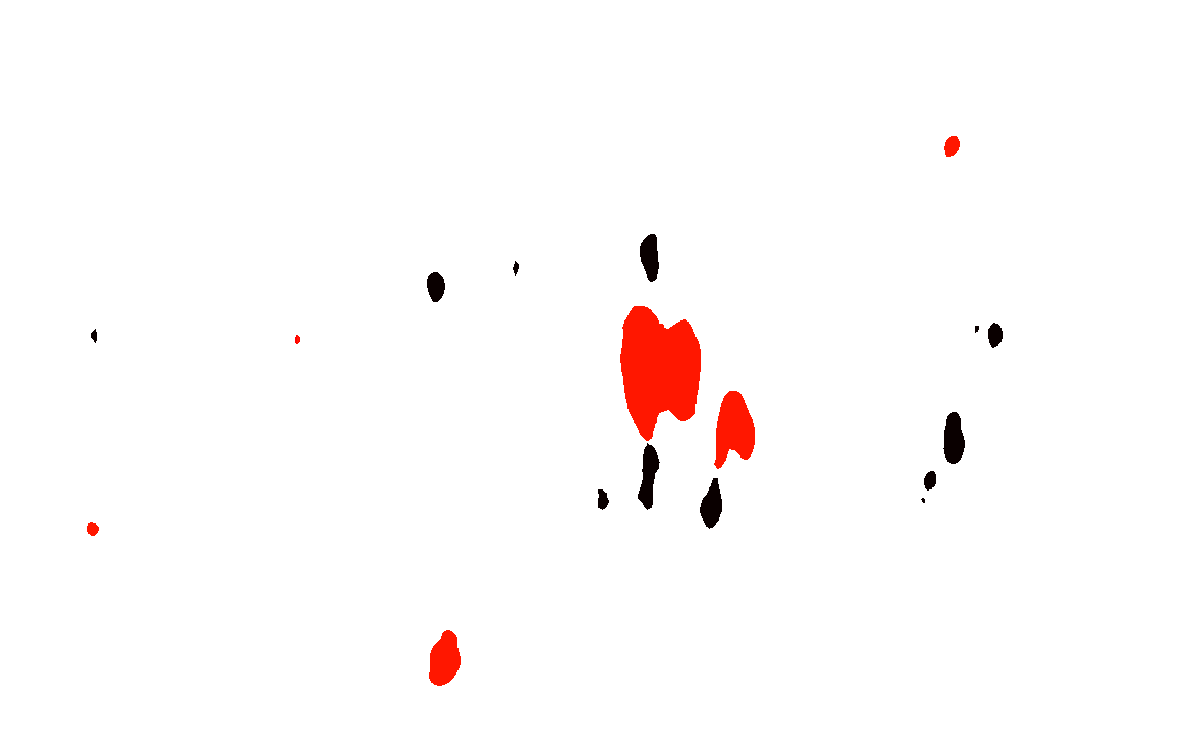}
\coordinate (ymax) at (0,10);
\coordinate (xmax) at (16, 0);
\coordinate (1) at (0,0);

\draw [thick] (0,0) -- (ymax);
\draw [thick] (0,0) -- (xmax);
\draw [thick] (16,10) -- (xmax);
\draw [thick] (16,10) -- (ymax);

\node [left of = 1, xshift = 0.3 cm]{\LARGE $-5$};
\node [left of = 1, yshift = 2 cm, xshift = 0.3 cm] {\LARGE $-3$};
\node [left of = 1, yshift = 4 cm, xshift = 0.3 cm] {\LARGE $-1$};
\node [left of = 1, yshift = 6 cm, xshift = 0.5 cm] {\LARGE $1$};
\node [left of = 1, yshift = 8 cm, xshift = 0.5 cm] {\LARGE $3$};
\node [left of = 1, yshift = 10 cm, xshift = 0.5 cm] {\LARGE $5$};
\node [left of = 1, yshift = 5 cm, xshift = -0.4cm] {\rotatebox{90} {\LARGE y[m]}};

\draw ($(1) + (-0.2,0)$) -- ++ (0.2,0);
\draw ($(1) + (-0.2,2)$) -- ++ (0.2,0);
\draw ($(1) + (-0.2,4)$) -- ++ (0.2,0);
\draw ($(1) + (-0.2,6)$) -- ++ (0.2,0);
\draw ($(1) + (-0.2,8)$) -- ++ (0.2,0);
\draw ($(1) + (-0.2,10)$) -- ++ (0.2,0);

\node [below of = 1, xshift = 0 cm, yshift = 0.25cm]  {\LARGE$0$};
\node [below of = 1, xshift = 2 cm, yshift = 0.25cm]  {\LARGE$2$};
\node [below of = 1, xshift = 4 cm, yshift = 0.25cm]  {\LARGE$4$};
\node [below of = 1, xshift = 6 cm, yshift = 0.25cm]  {\LARGE$6$};
\node [below of = 1, xshift = 8 cm, yshift = 0.25cm]  {\LARGE$8$};
\node [below of = 1, xshift = 10 cm, yshift = 0.25cm]  {\LARGE$10$};
\node [below of = 1, xshift = 12 cm, yshift = 0.25cm]  {\LARGE$12$};
\node [below of = 1, xshift = 14 cm, yshift = 0.25cm]  {\LARGE$14$};
\node [below of = 1, xshift = 16 cm, yshift = 0.25cm]  {\LARGE$16$};
\node [below of = 1, xshift = 8 cm, yshift = -0.7cm]   {\LARGE x[m]};

\draw ($(1) + (0,-0.2)$) -- ++ (0,0.2);
\draw ($(1) + (2,-0.2)$) -- ++ (0,0.2);
\draw ($(1) + (4,-0.2)$) -- ++ (0,0.2);
\draw ($(1) + (6,-0.2)$) -- ++ (0,0.2);
\draw ($(1) + (8,-0.2)$) -- ++ (0,0.2);
\draw ($(1) + (10,-0.2)$) -- ++ (0,0.2);
\draw ($(1) + (12,-0.2)$) -- ++ (0,0.2);
\draw ($(1) + (14,-0.2)$) -- ++ (0,0.2);
\draw ($(1) + (16,-0.2)$) -- ++ (0,0.2);

\node at (1, 3) {\LARGE\color{blue}{$\times^1$}};
\node at (4, 5.5) {\LARGE\color{blue}{$\times^2$}};
\node at (6, 1) {\LARGE\color{blue}{$\times^3$}};
\node at (8.5, 8.2) {\LARGE\color{blue}{$\times^4$}};
\node at (9, 5) {\LARGE\color{blue}{$\times^5$}};
\node at (10, 4) {\LARGE\color{blue}{$\times^6$}};
\node at (11.5, 2) {\LARGE\color{blue}{$\times^7$}};
\node at (13, 8.2) {\LARGE\color{blue}{$\times^8$}};
\node at (15, 4) {\LARGE\color{blue}{$\times^9$}};

\end{overpic}


\end{tikzpicture}}
			\caption{Robust LRT using outlier model}
			\label{fig_rob2_fullmiss}
		\end{minipage}
	\end{subfigure}
	\caption{Detection results for $\alpha = 0.05$ in high surface roughness environments. Color coding: detected targets (red), false alarms (black).}
	\label{fig_fullmiss}
\end{figure}
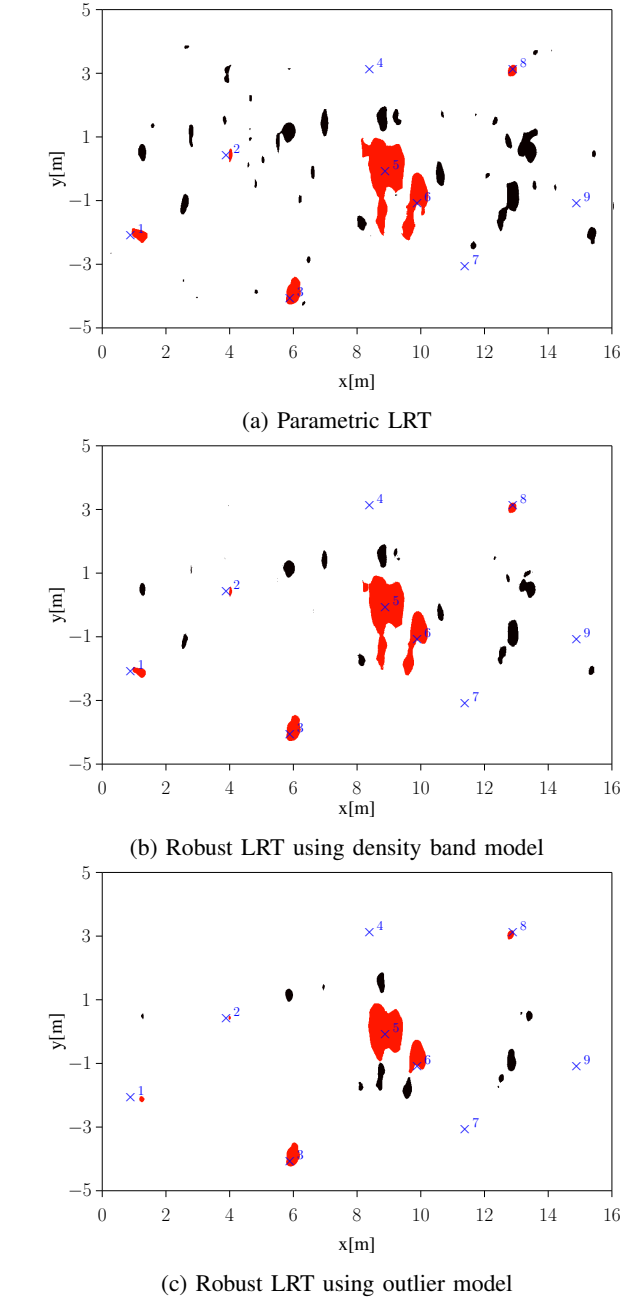

Finally, we consider a scenario where the entire investigation area admits a high roughness profile. Fig.~\ref{fig_fullmiss} shows the detection results for this scenario. Each detector commonly presents three missing targets, which are buried landmines number 4 and 9, and shell number 7. The increased number of errors is to be expected since this is an effect of a mismatch condition between the true distribution of the measurement and the assumed model. However, in comparison to the parametric model, the robust detectors are preferable with fewer false alarms. 

This detection performance can be improved by carefully assigning the band interval which naturally aligns with the non-stationary behavior of the measurements \cite{Pambudi2018, Zoubir2004}. A different approach can also be applied by considering the LRT which takes into account statistical dependencies between measurements \cite{Iyengar2011}. These methods are currently being explored and for a separate publication. The robust LRT using the 

Table~\ref{result} summarizes the performance of each detector in different rough surface environments. Accommodating uncertainties in the distribution with the robust LRTs reduces the number of false alarms significantly for all surface roughness levels. The robust LRT using the outlier model is preferable with only one missed detection for both low and mixed roughness profiles, with the lowest number of false alarms among all detectors. 

\begin{table}[t]
	\renewcommand{\arraystretch}{1.2}
	\caption{Number of false alarms (FA) and number of missed detections (MD) for different surface roughness.}
	\label{result}
	\centering
	\begin{tabular}{| l | c | c | c | c | c | c |}
		\hline
		\multirow{2}{2.3cm}{\textbf {LRT detector}}& \multicolumn{2}{c |}{\textbf {Low}} & \multicolumn{2}{c |}{\textbf{Mixed}} &  \multicolumn{2}{c|}{\textbf{High}} \\ \cline{2-7}
		& FA & MD &  FA & MD & FA & MD\\ \hline \hline
		Parametric model & \cellcolor{g}22 & 0 & \cellcolor{g}40 & 1 & \cellcolor{g}41 & 3\\ \cline{1-7}
		Robust (band model) & \cellcolor{g}12 & 1 & \cellcolor{g}19 & 1 & \cellcolor{g}17 & 3 \\ \cline{1-7}
		Robust (outlier model) &  \cellcolor{g}4 &1 & \cellcolor{g}8 & 1 & \cellcolor{g}12 & 3\\
		\hline
	\end{tabular}
\end{table}

\section{Conclusion}
\label{sec:conclusion}

We have investigated the problem of detecting landmines and unexploded ordnance in a rough surface environment using Forward-Looking Ground-Penetrating Radar. A minimax robust likelihood-ratio test has been designed by constructing a density band under each hypothesis and then finding the corresponding least favorable densities. The detection performance of the proposed robust detector has been evaluated using electromagnetic modeled data for different surface roughness and has been compared to alternative parametric approaches. Robust detectors have been shown to significantly reduce the false alarm rate while maintaining a comparable detection rate. However, it has also been shown that the uncertainty model needs to be chosen carefully in order to avoid potential over-robustification. All in all, robust hypothesis testing is a promising method to improve the accuracy of GPR landmine detectors that has not been sufficiently explored yet. In particular, the question of how uncertainty sets can be constructed in a data-driven manner certainly warrants further research.

\section*{Acknowledgment}

The authors would like to thank Dr. Traian Dogaru of US Army Research Laboratory for providing the simulated data.

\bibliographystyle{IEEEtran}
\bibliography{ref}

\vspace{-1.2cm}
\begin{IEEEbiography}[{\includegraphics[width=0.9in, height=1.15in, clip, keepaspectratio] {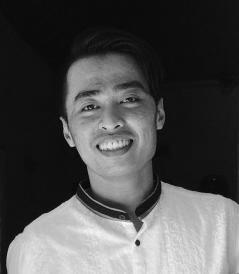}}] {Afief D. Pambudi} (S'16) is a member of the Signal Processing Group and an associate member of the Graduate School Computational Engineering, Technische Universi{\"a}t Darmstadt, Germany. Currently, he is working towards the PhD degree under the supervision of Prof. Abdelhak M. Zoubir. His research interests center around  topics  from  signal detection and estimation, robust statistics, and radar signal processing. 
\end{IEEEbiography}
\vspace{-1.2cm}
\begin{IEEEbiography}[{\includegraphics[width=0.9in, height=1.15in, clip, keepaspectratio] {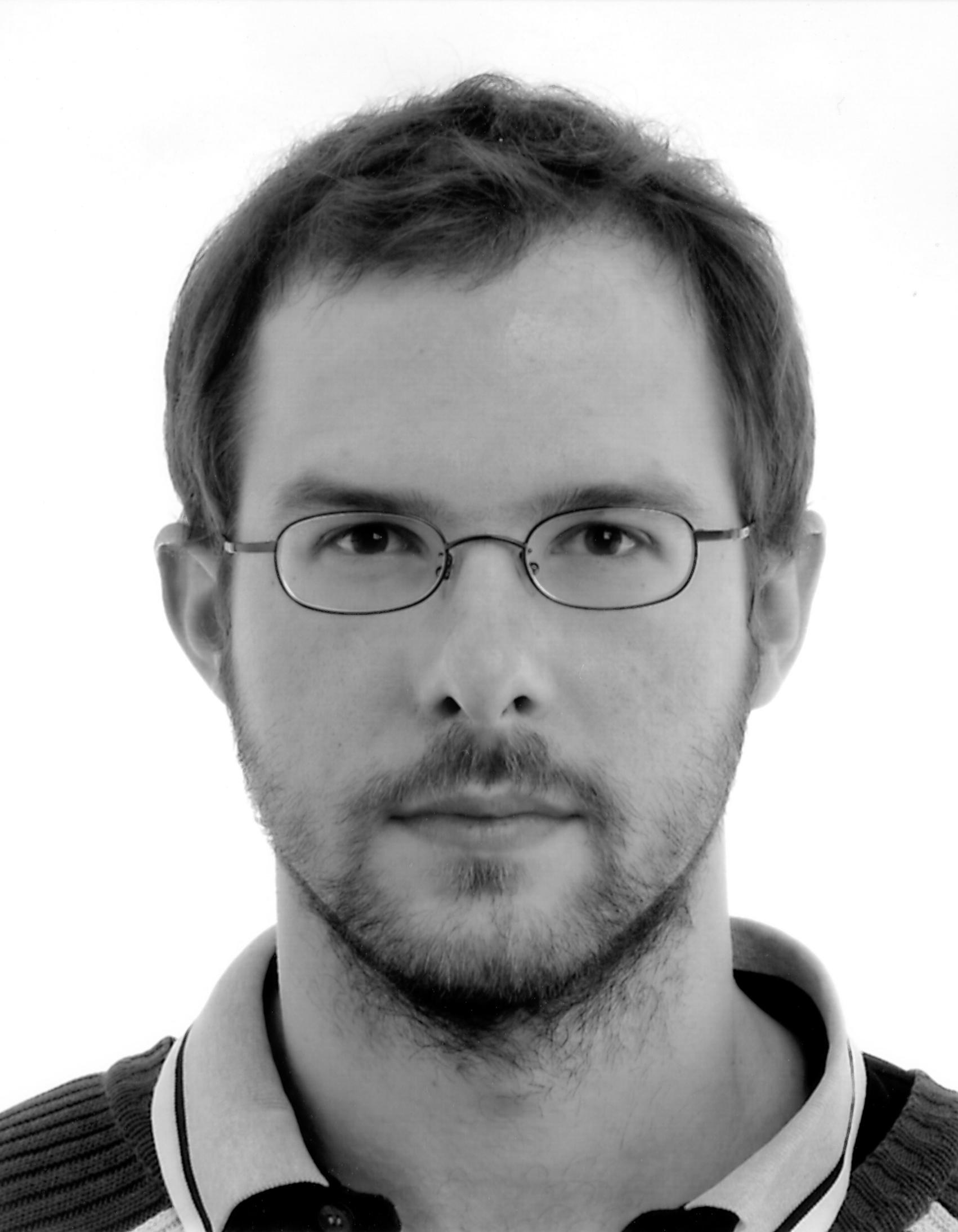}}] {Michael~Fau\ss{}} (M'17) received the Dipl.-Ing.~degree from Technische Universit\"at M\"unchen, Germany, in 2010 and the Dr.-Ing.~degree from Technische Universit\"at Darmstadt, Germany, in 2016, both in electrical engineering. In November 2011, he joined the Signal Processing Group at Technische Universit\"at Darmstadt and in 2017 he received the dissertation award of the German Information Technology Society for his PhD thesis on robust sequential detection. In September 2019 he joined Prof.~H.~Vincent Poor's group at Princeton University as a postdoc on a research grant by the German Research Foundation (DFG). His current research interests include statistical robustness, sequential detection and estimation, and the role of similarity measures in statistical inference.
\end{IEEEbiography}
\vspace{-1.2cm}
\begin{IEEEbiography}[{\includegraphics[width=0.9in, height=1.15in, clip, keepaspectratio] {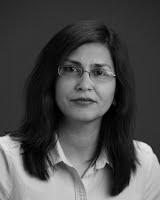}}] {Fauzia~Ahmad} (F'18) received her Ph.D. degree in electrical engineering from the University of Pennsylvania, Philadelphia, PA, USA, in 1997. Since 2016, she has been with Temple University, Philadelphia, PA, USA, where she is currently an Associate Professor with the Department of Electrical and Computer Engineering. Her research interests include the areas of statistical signal and array processing, computational imaging, compressive sensing, waveform design, target localization and tracking. She has authored over 240 journal articles and peer-reviewed conference papers and eight book chapters in the aforementioned areas. She is a Fellow of the International Society for Optics and Photonics (SPIE). She is a member of the Sensor Array and Multichannel Technical Committee of the IEEE Signal Processing Society,  the Computational Imaging Technical Committee of the IEEE Signal Processing Society, the Radar Systems Panel of the IEEE Aerospace and Electronic Systems Society, and the Electrical Cluster of the Franklin Institute Committee on Science and the Arts. She is an Associate Editor of the IEEE TRANSACTIONS ON COMPUTATIONAL IMAGING, and the IEEE TRANSACTIONS ON AEROSPACE AND ELECTRONIC SYSTEMS. She also serves on the editorial board of the IET Radar, Sonar, \& Navigation. She is the chair of the IEEE Dennis J. Picard Medal for Radar Technologies and Applications Committee.
\end{IEEEbiography}
\vspace{-1cm}
\begin{IEEEbiography}[{\includegraphics[width=0.9in, height=1.15in, clip, keepaspectratio] {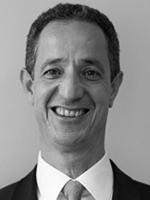}}] {Abdelhak~M.~Zoubir} (F'08) received the Dr.-Ing. degree from Ruhr-Universit{\"a}t Bochum, Germany, in 1992. He was with Queensland University of Technology, Australia from 1992 to 1998, where he was an Associate Professor. In 1999, he joined Curtin University of Technology, Australia, as a Professor of telecommunications. In 2003, he moved to Technische Universit{\"a}t Darmstadt, Germany as a Professor of signal processing and the head of the Signal Processing Group. His research interests include statistical methods for signal processing with emphasis on bootstrap techniques, robust detection and estimation and array processing applied to telecommunications, radar, sonar, automotive monitoring and safety, and biomedicine. He published more than 400 journal and conference papers on these areas. He is an IEEE Distinguished Lecturer (Class 2010-2011). He served as a General Chair and Technical Chair of numerous international IEEE conferences. He also served on publication boards of various journals, most notably, he served a three-year term as the Editor-In-Chief of the IEEE Signal Processing Magazine (2012-2014). He was the Chair (2010-2011), a Vice-Chair (2008-2009), and a Member (2002-2007) of the IEEE SPS Technical Committee Signal Processing Theory and Methods (SPTM). He was a Member of the IEEE SPS Technical Committee Sensor Array and Multichannel Signal Processing (SAM) from 2007 until 2012. He also served on the Board of Directors of the European Association of Signal Processing (EURASIP) (2009-2016) and on the Board of Governors of the IEEE SPS (2015-2017). He served as the President of EURASIP from 2017 to 2018.
\end{IEEEbiography}
\end{document}